\documentstyle[aps,psfig,epsfig]{revtex}
\begin{document}
\title{Statistical mechanics of a correlated energy landscape model 
for protein folding funnels}
\author{Steven S. Plotkin, Jin Wang, Peter G. Wolynes \\ 
Department of Physics and School of Chemical Sciences \\ 
University of Illinois \\
Urbana, IL 61801}

\maketitle

\begin{abstract}
	Energetic correlations 
	due to polymeric constraints and the locality of 
	interactions, in conjunction with the apriori 
	specification of the existence 
 	of  a particularly low energy state,
	provides a method of introducing the aspect of minimal frustration
	to the energy landscapes of random heteropolymers.  
	The resulting funnelled landscape exhibits both  a 
	phase transition from a molten globule to a folded state, and
	the heteropolymeric glass transition in the globular state. 
	We model the folding  transition
	in the self-averaging regime, which together with a simple theory of 
	collapse allows us to depict folding as a double-well
	free energy surface in terms of suitable reaction coordinates.  
	Observed trends in barrier positions and heights 
	with protein sequence length, stability, and 
	temperature are explained within
	the context of the model.  We also discuss the new 
	physics which arises from the introduction of explicitly cooperative 
	many-body interactions, as might arise from side-chain packing and 
	non-additive hydrophobic forces. Denaturation curves 
	similar to those seen in simulations are 
	predicted from the model.
\end{abstract}

\newpage



\def \bgeq{\begin{equation}}
\def \endeq{\end{equation}}
\def \zn{z_{\mbox{\tiny N}}}
\def \bgeqarray{\begin{eqnarray}}
\def \endeqarray{\end{eqnarray}}
\def \zbar{\overline{z}}
\def \Zbar{\overline{Z}}
\def \znq{z_{\mbox{\tiny NQ}}}
\def \delen{\delta \varepsilon_{\mbox{\scriptsize n}}}
\def \kboltz{k_{\mbox{\tiny B}}}
\def \tf{T_{\mbox{\tiny F}}}
\def \tg{T_{\mbox{\tiny G}}}
\def \tc{T_{\mbox{\tiny C}}}
\def \etaH{\eta_{\mbox{\tiny H}}}
\def \etaC{\eta_{\mbox{\tiny C}}}
\def \znc{z_{\mbox{\tiny N}_{\! c}}}
\def \znh{z_{\mbox{\tiny N}_{\! h}}}
\def \zh{z_{\mbox{\tiny N$-$NQ}}}
\def \NH{N_{\mbox{\tiny H}}}
\def \NC{N_{\mbox{\tiny C}}}
\def \etaHqz{\etaH\left( Q,\zbar \right)}
\def \ebar{\overline{\varepsilon}}
\def \qmg{Q_{\mbox{\tiny MG}}}
\def \zmg{\zbar_{\mbox{\tiny MG}}}
\def \qf{Q_{\mbox{\tiny F}}}
\def \qu{Q_{\mbox{\tiny U}}}
\def \zf{\zbar_{\mbox{\tiny F}}}
\def \qst{Q^{\star}}
\def \zst{\zbar^{\star}}
\def \delF{\Delta F^{\star}}
\def \gtrsim{ \leavevmode{\raisebox{-.5ex}{ $\stackrel{>}{\sim}$ } } }
\def \lessim{ \leavevmode{\raisebox{-.5ex}{ $\stackrel{<}{\sim}$ } } }
\def \Evect{\left( \varepsilon ,\ebar , \delen , \tf \right)}
\def \slev{s_{\mbox{\tiny LEV}}}

\renewcommand{\theequation}{\thesection.\arabic{equation}}

\section{Introduction}

Molecular scientists view protein folding as a complex chemical reaction.
Another fruitful analogy from statistical physics is that folding resembles a
phase transition in a finite system.  A new view of the folding process 
combines these two ideas along with the notion that a statistical characterization
of the numerous possible protein configurations is sufficient for understanding
folding kinetics in many regimes.

The resulting energy landscape theory of folding acknowledges that the energy
surface of a protein is rough, containing many local minima like the landscape 
of a spin glass.  On the other hand, in order to fold rapidly to a stable 
structure there must also be guiding forces that stabilize the native structure
substantially more than other local minima on the landscape.
This is the principle of minimum frustration~\cite{BW87}.
The energy landscape can be said then to resemble a ``funnel''.~\cite{BOSW}
Folding rates then depend on the statistics of the energy states as they become
more similar to the native state at the bottom of the funnel.

One powerful way of investigating protein energy landscapes has been the simulation of ``minimalist'' models.  These models are not fully atomistic, but caricature the protein as a series of beads on a chain either embedded in a continuum~\cite{friedrichs} or on a lattice~\cite{SO}.  A correspondence, in the sense of phase transition theory, between these models and real proteins has been set up using energy landscape ideas~\cite{owls}. Many issues remain to be settled however in understanding how these model landscapes and folding mechanisms change as the system under study becomes larger and as one introduces greater complexity into the modelling of this correspondence, as for example, by explicitly incorporating many-body forces and extra degrees of freedom.  Simulations become cumbersome for such surveys, and an analytical understanding is desirable.

Analytical approaches to the energy landscape of proteins have used much of the mathematical techniques used to treat spin glasses~\cite{mezard} and regular magnetic 
systems~\cite{landau}.  The polymeric nature of the problem must also be taken into account.  Mean field theories based on replica techniques~\cite{shak} and variational methods~\cite{sasai} have been very useful, but are more difficult to make physically intuitive than the straightforward approach of the random energy model~\cite{derridarem}, which flexibly takes into account many of the types of partial order expected in biopolymers~\cite{ZW}.
Recently we have generalized the latter approach to take into account correlations in the 
landscape of finite-sized random heteropolymers~\cite{pww}. 
This treatment used the formalism of the generalized random 
energy model (GREM) analyzed by
Derrida and Gardner~\cite{derridagardner}.  In this paper, we extend that analysis to take into account the minimum frustration principle and thereby treat protein-like, partially non-random heteropolymers.

There are various ways of introducing the aspect of minimum frustration to analytical models with rugged landscapes.
One way recognizes that many empirical potentials actually are obtained by a statistical analysis of a database, and when the database is finite, there is automatically an aspect of minimal frustration for any member of that database.  Thus the so-called ``associative memory'' hamiltonian models~\cite{friedrichs2} have co-existing funnel-like and rugged features in their landscape.  Other methods of introducing minimal frustration model the process of evolution as giving a Boltzmann distribution over sequences for an energy gap between a fixed target structure and unrelated ones~\cite{shaknature}. All of the above approaches can be straightforwardly handled with replica-based analyses.  Here we show that the GREM analyses can be applied to minimally frustrated systems merely by requiring the energy of a given state to be specified 
as having a particularly low value.  Minimally frustrated, funnelled landscapes are just a special case of the general correlated landscape studied earlier.

A convenient aspect of the correlated landscape model is that it allows the treatment of the polymer physics in a very direct way, using simple statistical thermodynamics in the tradition of Flory~\cite{flory}.  Here we will show how the interplay of collapse and topological ordering can be studied.  In order to do this we introduce a simple ``core-halo'' model to take into account the spatially inhomogenous density.  We will also discuss the role of many-body forces in folding.  Explicitly cooperative many-body forces have often been involved in the thinking about protein structure formation.  Hydrophobic forces are often modelled as involving buried surface area.  Such an energy term is not pairwise-additive but involves three or more interacting bodies.  Side-chain packing involves objects fitting into holes created by more than one other part of the chain, thus the elimination of side-chains from the model can yield an energy function for backbone units with explicit non-additivity.  These many-body forces can be treated quite easily by the GREM, and we will see that they can make qualitative changes in the funnel topography.

To illustrate the methods here, we construct two-dimensional free energy surfaces for the folding funnel of minimally frustrated polymers.  These explicitly show the coupling between between density and topological similarity in folding.  We pay special attention to the location of the transition state ensemble and discuss how this varies with system size, cooperativity of interactions, and thermodynamic conditions.  In the case of the $27$-mer on a lattice, a detailed fit to the lattice simulation data~\cite{SO} is possible.  Although delicate cancellations of energetic and entropic terms are involved in the overall free energy, plausible parameters fit the data.

The trends we see in the present calculations are very much in harmony with the experimental information on the nature and location of the transition state ensemble~\cite{oas,fersht}.  We intend later to return to this comparison, especially taking into account more structural details within the protein.

The organization of this paper is as follows: In Sec. 2 we introduce a theory of the free energy at constant density, and in this context investigate the effects of cooperative interactions on the transition state ensemble and corresponding free energy barrier. In Sec. 3 we detail a simple theory coupling collapse 
with topological similarity, and resulting in the ``core-halo'' model described there.  In Sec. 4 we apply this collapse theory to obtain the free energy in terms of density and topological order, now coupled via the core-halo model.  In the same section we compare our model of the minimally frustrated heteropolymer with lattice simulations of the $27$-mer.  
In terms of the categorization of Bryngelson et. al.~\cite{BOSW} these free energy surfaces depict scenarios described as type I or type IIa folding.
We then study the quantitative aspects of the barrier as a function of the magnitude of $3$-body effects.  The dependence of position and height of the barrier as a function of sequence length is studied, as well as the effects of increasing the stability gap.  Finally, we study the denaturation curve as determined by the constant and variable density models.
In Sec. 6 we discuss the results and conclude with some remarks.

\section{A Theory of the Free Energy}
\setcounter{equation}{0}

In this section, we show how the apriori specification of the existence of 
a particularly low energy configuration,
together with a theory of energy correlations for configurationally 
similar states, leads to a model for the folding transition and 
corresponding free energy surface in 
protein-like heteroplolymers.
This ansatz for the correlated energy landscape corresponds to the 
introduction of minimal frustration in a random energy landscape, where
the order parameter here (which will function as a reaction coordinate
for the folding transition) counts the number of native contacts
or hydrogen bonds.

The GREM theory for random heteropolymers developed by us earlier
investigates the 
interplay between entropy loss and energetic roughness as a function of 
similarity to any given reference state, all at fixed density. However
for exceptional reference states such as the ground state of a well-packed
protein, the density
is not independent of configurational similarity, so a theory of the 
coupling of density with topological similarity must also be developed (section 3).

We start by assuming a simple ``ball and chain'' model for a protein
which is readily comparable with simulations,
for example of the $27$-mer, which is widely believed to capture many of the 
quantitative aspects of folding (section 4).  
Proteins with significant secondary structure have an effectively reduced number of 
interacting units as may be described by a ball and chain model.
Properties of both, when appropriately scaled by 
critical state variables such as the folding temperature $\tf$, glass temperature
$\tg$, and collapse temperature $\tc$,
will obey a law of corresponding states~\cite{owls}.
Thus the behavior describing a complicated real protein can be validly described
by an order parameter applied to a 
minimal ball and chain model in the same universality class.
For a $27$-mer on a 3-dimensional cubic lattice, there are 
$28$ contacts in the most collapsed (cubic) structure.  For concreteness
 we take such a maximally compact structure to be 
the configuration of our ground state, the generalization to a less compact
ground state being straightforward in the context of the model to be described. 
For a collapsed polymer of sequence length $N$,
the number of pair contacts per monomer , $\zn$,
is a combination of a bulk term, a surface term,
and a lattice correction~\cite{DI}.  The effect of the surface on the number of 
contacts is quite important even for large macromolecules, as $ z_3(N)$ approaches 
its bulk value of 2 contacts per monomer rather slowly, as 
$ \sim 2 - 3 N^{-1/3} $.

To describe states that are not completely collapsed, we introduce the packing 
fraction $\eta \cong N \sigma/R_g^3$ as a measure of the density of the 
polymer, where $\sigma$ is the volume per monomer
and $R_g$ is the radius of gyration of the whole protein.  So for less dense 
states the total number of contacts is reduced from its collapsed value 
$N\zn$, to $N\zn\eta$.

In the spirit of the lattice model we have in mind for concreteness,
we introduce a simple contact hamiltonian to determine the energy of 
the system:
\bgeq
	{\cal H} = \sum_{i<j} \varepsilon_{ij} \sigma_{ij}  \; ,
\label{eq:ham}
\endeq
where $\sigma_{ij} = 1 $  when there is a 
contact made between monomers $\left\{ i j \right\}$ in the chain, and 
$\sigma_{ij} = 0 $ otherwise. 
Here contact means that two monomers $\left\{ i j \right\}$, 
non-consecutive in sequence along the backbone chain, are 
adjacent in space at neighboring lattice sites.
$\varepsilon_{ij}$ is a random variable so that, at constant
density, the total energies of the various configurations, each 
the sum of many $\varepsilon_{ij}$, are approximately gaussianly 
distributed by the central limit theorem, with mean energy 
(at a given density $\eta$) $\overline{E_\eta} = 
N \zn \eta \overline{\varepsilon}$,
(where $\overline{\varepsilon}$ is simply defined as the mean energy per
contact and $N \zn \eta$ is again the total number of contacts),
and variance $\Delta \! E^2_{\eta} = N \zn \eta \varepsilon^{2}$,
where $\varepsilon^{2}$ is the effective width of the energy 
distribution per contact.

Suppose there exists a configurational state $n$ of energy $E_n$ 
(which will later become the ``native'' state).
Then if the Hamiltonian for our system is defined as in eq. (\ref{eq:ham}),
we can find the probability that configuration $a$ has energy $E_a$,
given that $a$ has an overlap $Q_{an}$ with $n$,~\cite{pww}
where $Q_{an} \equiv Q$ is
the number of contacts that state $a$ has in common
with $n$, divided by the total number of contacts $N\zn\eta$ (since 
this analysis is at constant density both $a$ and $n$ have 
$N\zn\eta$ contacts).  This distribution is simply a gaussian with a 
$Q$ dependent mean and variance:
\bgeq
	\frac{P_{an}\left(E_a|Q|E_n\right)}{P_n\left(E_n\right)} \sim
	\exp\left( -\frac{\left[\left(E_a-\overline{E}\right) - Q\left(
	E_n-\overline{E}\right)\right]^{2}}{2 N \zn \eta \varepsilon^2
	\left(1-Q^2\right)}\right)
\label{eq:jointP}
\endeq
When $Q=1$ states $a$ and $n$ are identical and must then have the 
same energy, which (\ref{eq:jointP}) imposes by becoming delta function,
and when $Q=0$ states $a$ and $n$ are uncorrelated and then
(\ref{eq:jointP}) becomes the gaussian distribution of the Random 
Energy Model for the energy of state $a$~\cite{BW87}.
Expression (\ref{eq:jointP}) holds for all states of the same 
density as $n$, e.g. all collapsed states if $n$ is the native state
(the degree of collapse must be a somewhat coarse-grained 
description to avoid fluctuations due to lattice effects 
coupled with finite size).

Previously a theory was developed of the configurational
entropy $S_{\eta}\left(Q\right)$ 
as a function of similarity $Q$, at constant
density $\eta$.~\cite{pww} Given $S_{\eta}\left(Q\right)$
and the conditional probability distribution (\ref{eq:jointP}),
the average number of states of energy $E$ and overlap $Q$ 
with state $n$, all at density $\eta$, is then
\bgeqarray
	\left< \mbox{n}_{\eta}\left(E|Q|E_n\right) \right> &=& 
	\mbox{\large e}^{S_{\eta}\left(Q\right)} \times
	\frac{P\left(E|Q|E_n\right)}{P\left(E_n\right)}
	\nonumber \\
	&\sim& \exp N \left\{ s_{\eta}\left(Q\right) -
	\frac{1}{2\left(1-Q^2\right)}\left(\frac{
  	\left(E-\overline{E}\right) - Q\left(
	E_n-\overline{E}\right)}{NJ_{\eta}}\right)^2 \right\}
\label{eq:ne}
\endeqarray
where $J_{\eta}^2 \equiv \zn \eta \varepsilon^2$ and 
$s_{\eta}\left(Q\right) \equiv S_{\eta}\left(Q\right)/N$~\cite{snote}. 
Equation (\ref{eq:ne})
is still gaussian with a large number of states
provided $E < E_c$ (for negative energies) where $E_c
= Q E_n + N J_{\eta}\sqrt{2\left(1-Q^2\right)s_{\eta}\left(Q\right)}$
is a critical energy below which the exponent changes sign and 
the number of states becomes
negligably small.

At temperature $T$, the Boltzmann factor 
$\frac{1}{Z}e^{-E/T}$~\cite{kboltznote}
weighting each state shifts the number distribution of energies
so that the maximum of the thermally weighted distribution 
can be interpreted as the most 
probable (thermodynamic) energy at that temperature
\bgeq
E_{\eta}\left(T,Q,E_n\right) = \overline{E} + 
	Q\left(E_n - \overline{E}\right) - 
	\frac{N \zn \eta \varepsilon^2 \left(1 - Q^2\right)}{T} \; .
\label{eq:et}
\endeq
The above expression for the most probable energy is useful
provided the distribution (\ref{eq:ne}) is a good measure of the 
actual number of states at $E$ and $Q$, the condition for which is 
that the fluctuations in the number of states be much smaller than 
the number of states itself. To this end, we make here the 
simplifying assumption that in each ``stratum'' defined by the 
set of states which have an overlap $Q$ with the native state,
the states themselves are not further correlated with each other, i.e.
$P\left(E_a,Q,E_b,Q | E_n\right) = P\left(E_a,Q,E_n\right)
P\left(E_b,Q,E_n\right)$, so that in each stratum of the 
reaction coordinate $Q$, the set of states is modelled by a 
random energy model.  Then since the number of states 
$\mbox{n}_{\eta}\left(E|Q|E_n\right)$ counts a collection of 
random uncorrelated variables, large when $E>E_c$, the relative 
fluctuations $\sqrt{\left<\left(n-\left< n\right>\right)^2\right>}/
\left< n\right>$ are $\sim \left< n\right>^{-1/2}$
and are thus negligible.  
So $n\left(E|Q|E_n\right) \approx \left< n\left(E|Q|E_n\right)\right>$,
and we can evaluate the exponent in the number of states (\ref{eq:ne}) 
at the the most probable energy (\ref{eq:et}) as an accurate measure
of the ($Q$ dependent) thermodynamic entropy at temperature $T$
\bgeq
S_{\eta}\left(T,Q,E_n\right) = S_{\eta}\left(Q\right) -
	\frac{N \zn \eta \varepsilon^2 \left(1 - Q^2\right)}{2 T^2} \; .
\label{eq:sqt}
\endeq

The assumption of a REM at each stratum of $Q$
is clearly a first approximation to a more accurate
correlational scheme.  
The generalization to treat each stratum itself as a GREM as in our 
earlier work is nevertheless straightforward, since our earlier work suggested 
only quantitative changes, which we will not pursue here.
If two configurations $a$ and $b$ have overlap
$Q$ with state $n$ and thus are correlated to $n$ energetically, 
they are certainly correlated to each other, particularly for 
large overlaps where the number of shared contacts is large. 
Using the REM scheme at each stratum is more accurate for small $Q$ 
and breaks down to some extent for large $Q$.
In the ultrametric scheme of the GREM, states $a$ and $b$ have an 
overlap $q_{ab} \geq Q$, which is more accurate for large overlap
than at small $Q$ where states $a$ and $b$ need not share {\em{any}} bonds
and still can both have overlap $Q$ with $n$.
One can also further correlate the energy landscape 
of states by stratifying with respect to $q_{ab} = q$ and so on,
resulting in a hierarchy of overlaps and correlations best treated 
using renormalization group ideas.

Just as the number of states (\ref{eq:ne}) has a characteristic energy 
for which it vanishes, the REM entropy for a stratum at $Q$
(\ref{eq:sqt}) vanishes at a 
characteristic temperature
\bgeq
\frac{T_g\left(Q\right)}{\varepsilon} = \sqrt{\frac{\zn\eta\left(
	1-Q^2\right)}{2s_{\eta}\left(Q\right)}}
\label{eq:tgs}
\endeq
which signals the trapping of the polymer into a low energy 
conformational state within the stratum characterized by $Q$.

If $T_g(Q)$ is a monotonically decreasing function of $Q$,
as the temperature is lowered the polymer will gradually be
thermodynamically confined in its conformational search to smaller 
and smaller basins of states.  The basin around the native state is
the largest basin with the lowest ground state, and hence is the first 
basin within which to be confined.  
Its characteristic size at temperature $T$ is just the number 
of states within overlap $Q_o(T)$, where  $Q_o(T)$ is the value of 
overlap $Q$ that gives $T_g(Q_o) = T$ in equation (\ref{eq:tgs}).
Thus there is now no longer a single glass temperature at
which ergodic confinement suddenly occurs, as in the REM,
but there is a continuum of basin sizes to be localized within at
corresponding glass temperatures for those basins.

If $T_g(Q)$ has a single maximum at say $Q^{\star}$, the 
glass transition is characterized by a sudden REM-like 
freezing to a basin of configurations whose size is determined 
by $Q^{\star}$.  The range of glass temperatures will turn out to be lower than
the temperature at which a folding transition occurs (see Fig.~\ref{fig:tgQZ}),
so that this model predicts a protein-like heteropolymer
whose folded state is stable by several $k_{\mbox{\tiny B}}T$ at 
temperatures where freezing becomes important.  A replica-symmetric
analysis of the free energy is therefore sufficient to describe the 
folding transition to such deep native states that are minimally 
frustrated.

From the thermodynamic expressions for the energy (\ref{eq:et}) and 
entropy (\ref{eq:sqt}) with the mean energy at density $\eta$, we can write 
down the free energy per monomer above the glass temperature
as the sum of 4 terms
\bgeq
\frac{F_{\eta}}{N}\left(T,Q,E_n\right) =
	\zn \eta \overline{\varepsilon} + Q \zn \delen
	- T s_{\eta}\left( Q\right) - 
	\frac{\zn \eta \varepsilon^2}{2 T} \left(1-Q^2\right) \; ,
\label{eq:fqt}
\endeq
where $\zn \delta \varepsilon_{n} = \zn (\varepsilon_n - 
\overline{\varepsilon}) = \left(E_n - \overline{E}\right)/N$ is the 
extra energy for each bond beyond the mean homopolymeric attraction energy
(the energy ``gap'' between an average molten globule structure and 
the minimally frustrated one),
times the number of bonds per monomer, 
and $s_{\eta}\left( Q\right) = S_{\eta}\left( Q\right)/N$ is the 
entropy per monomer.

The first term in (\ref{eq:fqt}) multiplied by $N$ is just 
the homopolymeric attraction energy between all the monomers,
for a polymer of density $\eta$.  It depends only on the degree of 
collapse, and  not on how many contacts are native contacts.
The second term is the average extra bias energy if a contact
is native, times the average number of native contacts per 
monomer.  The third term measures the equilibrium bias 
towards larger configurational entropy at smaller values of the 
reaction coordintate $Q$.
The last term accounts for the diversity of energy states that exist
on a rough energy landscape, the variance of which lowers 
thermodynamically the energy more than the entropy, and so
lowers the equilibrium free energy.

For a special surface in $\left(\delen,\varepsilon ,T\right)$ space,
expression (\ref{eq:fqt}) has a double minimum structure in the 
reaction coordinate $Q$, with one entropic minimum at low $Q$ 
corresponding to the ``molten globule'' state, separated by a barrier 
from an energetic 
minimum at high $Q$ corresponding to a ``folded'' state.
For a given temperature, values of $\delen$ and $\varepsilon^2$ 
can be obtained which
are reasonably close to the values obtained by a more accurate analysis
which includes the coupling of density with topology,
but we will not examine the constant density case in much detail for 
reasons discussed below,
except to remark that 1.) The true coupling between density and 
$Q$-constraints need not be strong to obtain a double-well free
energy structure, 2.) For monomeric units with pair interactions, the 
molten globule and folded minima are
{\em not} at $Q=0$ and $1$ respectively.  The position of the 
molten globule state is near the maximum of the entropy of the system,
which is at $Q \cong 0.1$ for the $27$-mer due to the interplay of
confinement effects and the combinatorial mixing entropy inherent 
in the ``coarse-grained'' description $Q$.~\cite{pww}
The native minimum shifts to $Q=1$ when many-body interactions 
are introduced (see the next section ).
3.) The barrier height,
at position $Q^{\circ} \cong 0.25$ for the $27$-mer with protein-like 
parameters ($\tf/\tg \cong 2$),
is small ($\Delta F^{\circ} \approx \kboltz \tf$), due to the 
effective cancellation of entropy loss by negative energy gain, as 
the system moves toward the native state (This cancellation is 
reduced when many-body forces are taken into account).
4.) When a linear form for the entropy is used in equation (\ref{eq:fqt}), 
e.g. $s(Q)=s_o(1-Q)$  instead of the more accurate $s(Q)$ obtained 
in reference \cite{pww},
the double minimum structure disappears and is replaced by a 
single minimum near $\tf$ at $Q \approx 1/2$, with the 
$Q=0$ and $Q=1$ states becoming free energy maxima. So folding is 
downhill or spinodal-like in this approximation.

\subsection{Effects of cooperative interactions}

As the interactions between monomeric segments become more explicitly cooperative,
the energetic correlations between states become significant only at 
greater similarity, with the system approaching the 
REM limit for $\infty$-body interactions, where the statistical 
energy landscape assumes a rough ``golf-course'' topography with a steep
funnel close to the native state.

In the presence of many-body interactions, the homopolymer collapse energy
also scales as a higher power of density ($\sim \ebar \, \zbar^{m-1}$).  
For even moderate $m \sim {\cal{O}}
\left( 1 \right)$ a first-order phase transition to collapsed states results,
which effectively confines all reaction paths in the coordinate $Q$ 
between molten globule and folded states to those where the density is 
constant and $\approx 1$. So within this constant density 
approximation we can investigate the 
nature of the folding transition as a function of the cooperativity of the
interactions, and see how the correlated 
landscape simplifies to the REM in the limit of $m$-body interactions with 
large $m$.

In the presence of $m$-body interactions, the $Q$ dependence in the
pair energy distribution (\ref{eq:jointP}) scales with $Q$ as $Q^{m-1}$.  Using 
this modified pair distribution along with the collapsed homopolymeric 
state as our zero point energy, the free energy (\ref{eq:fqt}) becomes
\bgeq
\frac{F}{N}\left(T,Q,E_n\right) = - T s_{1}\left( Q\right) - 
	Q^{m-1} \zn \left| \delen \right| -
	\frac{\zn \varepsilon^2}{2 T} \left(1-Q^{2\left( m-1\right)} \right) \; ,
\label{eq:fq1}
\endeq
where $s_{1}\left( Q\right)$ is the entropy as a function of constraint $Q$ 
for a fully collapsed polymer.
For pure $3$-body interactions and higher, the globule and folded 
states are very nearly at $Q \cong 0$ and $Q \cong 1$ respectively
(see figure~\ref{fig:FvsQ3and12}A). 
To the extent that this approximation is good, 
we can equate the free energies of the 
molten globule and folded structures
and obtain an $m$-independent
folding temperature (note again that this is not a good approximation for 
pair interactions):
\bgeq
T_{\mbox{\tiny F}} = \frac{\zn \left| \delen \right|}{2 s_o} \left(
	1 + \sqrt{ 1 - \frac{2 s_o \varepsilon^2}{\zn \delen^2}} \right)
\label{eq:tf}
\endeq
where $s_o$ is the maximum of the entropy as a function of constraint $Q$
(essentially the $\log$ of the total number of configurations).

From expression (\ref{eq:tf}) we can obtain a first approximation to
the constraint on the magnitude of the gap energy $\delen$ in order
to have a global folding transition (rather than merely a local glass 
transition) to the low energy state in question.  The condition that the
square root term in eq. (\ref{eq:tf}) be real gives the minimum 
gap for global foldability in terms of the roughness $\varepsilon$:
\bgeq
\frac{\delen^{\left(c\right)}}{\varepsilon} = \sqrt{\frac{2 s_o}{\zn}}
	\approx  \sqrt{2}
\label{eq:foldability}
\endeq
where the minimum folding temperature is then 
$\kboltz T_{\mbox{\tiny F}}^{\left(c\right)} \approx 
\delen^{\left(c\right)}/2$ (or equivalently, one can obtain the maximum
roughness for foldability as $\approx 1/\sqrt{2}$ of a given gap energy).
For typical proteins (with folding temperatures at $\approx 330 \, \mbox{K}$)
gap energies are (at least) $\approx 1 \, 
\mbox{kCal/mol}\cdot~\mbox{$\!\!$(lattice~unit)}$.
Note that eq. (\ref{eq:foldability}) is precisely the same result,
as it should be, to that obtained previously~\cite{goldstein} 
in the context of finding optimal folding energy functions, 
by requiring the quantity
$\tf/\tg > 1$, where the glass temperature $\tg = \sqrt{\zn \varepsilon^2 /
(2 s_o)}$ is evaluated at the molten globule overlap $Q = Q_g$.
From these arguments one can see that the distinction between the 
folding transition and the glass transition is a quantitative one 
characterized by the distinction between global and local basin sizes, but a 
crucial one for the sub-class of biological heteropolymers.

Evaluating $F\left(T,Q,E_n\right)/N$ (eq. (\ref{eq:fq1})) with 
protein-like energetic parameters at
the folding temperature $T_{\mbox{\tiny F}}$,
we obtain free energy curves as in Fig.~\ref{fig:FvsQ3and12}A,
plotted for example with $m=3$ and $m=12$, for a $27$-mer lattice protein.

Note that the transition state ensemble (the collection of states at 
$Q=Q^{\star}$, where $Q^{\star}$ is where the free energy is a 
maximum) becomes more and more native like (and thus the ensemble 
becomes smaller and smaller, eventually going to $1$ state in the REM)
as the energy correlations become more short-ranged in $Q$ (i.e. as $m$
increases) - see figure~\ref{fig:FvsQ3and12}B. The corresponding free energy 
barrier then grows with $m$  as the energetic 
bias ($\sim Q^{m-1}$) overcomes the entropic barrier only much closer
to the native state, and the barrier becomes
more and more entropic and less energetic (see fig.~\ref{fig:FvsQ3and12}C ). 

As was already mentioned, the above analysis was for a polymer 
of constant collapse density. However, numerical evidence for 
lattice models of protein-like heteropolymers 
suggests a coupling of density with nativeness,
with energetically favorable native-like states typically being denser. 
So to this end we now investigate in detail
a simple interpolative theory coupling collapse density $\eta$ with nativeness 
$Q$, assuming a native ($Q=1$) state which is completely collapsed ($\eta=1$).  
Including this effect in eq. (\ref{eq:fqt}) will complete our simple 
model of the folding funnel topography in two reaction-coordinate 
dimensions.

\section{A Theory of Collapse}
\setcounter{equation}{0}

At low degrees of nativeness, we expect that collapse should be roughly
homogeneous throughout the polymer, so that the density $\eta$ should 
depend only on the total number of contacts $\Zbar = N\zbar$, where 
$\zbar$ is the average number of contacts per monomer. This is the 
case if we straightforwardly apply $N\zn\eta=N\zbar$, so that 
$\eta=\zbar/\zn$, and any theory of $\eta$ must have this form in 
the low $Q$ limit (as well as $\eta \rightarrow 1$ as $Q \rightarrow 1$).
Now as we progress towards more native structures (higher $Q$), we 
should introduce a model that distinguishes between the constrained 
native structures or regions (in space) fixed or ``frozen'' by virtue 
of their native contacts, and those non-native regions, typically less dense,
and not constrained by any native contacts.  This model will impose a 
$Q$ dependence on $\eta$ by ascribing different densities to the 
collapsed, native ``core'' region(s) and the non-native, less dense 
``halo'' region(s). We adopt the simplest model that there is one native
core region of density $\eta\cong 1$, surrounded by a halo region with
$\etaH \leq 1$ (see fig.~\ref{fig:corehalo}). To find the $Q$ and $\zbar$
dependence of $\etaH$, we see that the total number of contacts $N\zbar$ is 
the sum of two terms
\bgeq
N\zbar =  \NC \znc + \NH \etaH \znh \; .
\endeq
The first term is the number of contacts inside the native core region,
where $\NC$ is the number of frozen monomers and $\znc$ is the number of 
contacts per monomer in the core. The number of contacts per monomer in 
a three dimensional collapsed walk of length $N$, mentioned above in Sec. II,
is given approximately by
\bgeq
\zn \approx \frac{1}{N} \mbox{Int}[2 N - 3 (N+1)^{2/3} + 3]
\label{eq:ZofN}
\end{equation}
where $\mbox{Int}[\ldots ]$ means take the integer part. So the number of 
contacts per monomer in the core is eq. (\ref{eq:ZofN}) with $N$ replaced by 
the number of core monomers, $\NC$. 
The second term is the total number of contacts in the halo, where $\NH$ is 
the number of monomers in the halo $N-\NC$, and $\etaH\znh$ is the 
approximate number of
contacts per monomer.  The packing fraction $\etaH$ can be interpreted as 
the probability that a monomer is within a region of space $\Delta\tau \cong
b^3$ where $b^3$ is the size of a monomer, which reduces the 
number of contacts per monomer from its collapsed value, $\znh$, to $\etaH\znh$.
$\znh$ is eq. (\ref{eq:ZofN}) with $N \rightarrow \NH$, which accounts for the fact 
that the halo has both an inner and outer surface.  Contacts at the interface of 
the core and halo are neglected.

Next we assume that basically all the {\em native} contacts are made in the dense
core, so that $Q$, the number of native contacts over the total number of 
possible native contacts, is given by $\NC\znc/N\zn$.
Then using $\NH=N-\NC \cong N - N\zn Q/\znq$, we can express
$\etaH$ as a function of $Q$ and $\zbar$
\bgeq
\etaH\left( Q,\zbar \right) = \frac{\zbar - \zn Q}{\zh\left(1-
	\frac{\zn}{\znq} Q \right)} \; .
\label{eq:etaH}
\endeq
The condition $\etaH > 0$ corresponds to the condition that the 
number of native bonds $N\zn Q$ cannot exceed the total number of 
bonds $N \zbar$. Note that $\etaH \rightarrow \zbar /\zn = \eta_{tot}$ at low
nativeness (small $Q$), where the polymer is almost all halo.
As the polymer becomes native-like ($Q \rightarrow 1$), contacts at the 
surface between the core and the halo become more important, and the simple
theory begins to break down. 

In the next section, we obtain the free energy in terms of the 
reaction coordinates $Q$ and $\zbar$ through the introduction of 
the halo density obtained above, but
we can now
re-investigate the glass transition temperature as a function 
of both $Q$ and $\zbar$ through the insertion of the halo density 
$\etaHqz$ into (\ref{eq:tgs}), giving the 
regions in the space of these reaction 
coordinates where the dynamics would tend to become glassy if $T_g(Q,\zbar)$
were comparable to $\tf$ (see fig. \ref{fig:tgQZ}). The values of 
$\tf/\tg \left(Q,\zbar\right)$ are always greater than $\cong 1.6$, 
as one can see from the figure, and so the assumption
of self-averaging used in section 2 is valid here, 
with the exception of very native-like 
states (at high $Q$ and $\zbar$), 
where the free energy becomes strongly sequence dependent for 
a finite size polymer.  

Of course $\tf/\tg \left(Q,\zbar\right)$ is a rather crude
measure of self-averaging, and a more rigorous method would be to 
follow the calculations by Derrida and Toulouse~\cite{DT} 
of the moments of the probability distribution
of $Y = \sum_j W_j^2$, 
measuring the sample to sample fluctuations of the sum of weights of 
the free energy valleys, and generalize them to finding the 
probability distribution of 
$Y(Q,\zbar)$.

The shape of the free energy surface is very sensitive to the 
form of $\etaHqz$, and this model of collapse represents one 
of the cruder approximations of the theory.  It predicts a weakly decreasing
halo density as $Q$ increases, and predicts at $\tf$ a folded state 
with a significant halo, that 
has overall less contacts than in the MG state (although there is 
of course a much larger native core).  This over-expansion of the halo
compensates entropically for the large loss in entropy due to the 
$N \zn Q$ constraints.

\section{The Density-Coupled Free Energy}
\setcounter{equation}{0}

The halo density $\etaHqz$ will appear in the roughness term of (\ref{eq:fqt}) 
since this term arises as a result of non-native interactions which contribute 
to the total variance of state energies.  
$\etaH$ would also appear in the 
entropic contribution in (\ref{eq:fqt}) because the parts of the polymer 
contributing to the entropy are the dangling loops or pieces constituting 
the halo - the frozen native core is fully constrained, but
slightly more accurate values are obtained specifically for the barrier
position $\qst$ if in the entropic term
we use an interpolation between the 
the low $Q$ formula of the density (to which it simplifies anyway at
weak topological constraint), and the 
core-halo formula (\ref{eq:etaH}) at high $Q$:
$$
\etaH^{tot} \left( Q,\zbar \right) = \left(1-Q\right) \frac{\zbar}{\zn}
	+ Q \frac{\zbar - \zn Q}{\zh\left(1-
	\frac{\zn}{\znq} Q \right)} \; .
$$
This gives more weight to the two behaviors in their respective regimes: 
mean-field uniform density at weak constraint or low $Q$, and core-halo behavior
at strong constraint. The pure halo density
formula (\ref{eq:etaH}) is still used in the roughness term (but this is not 
a crucial point).
The {\em total} density $\eta_{tot} = \zbar/\zn$ appears in the homopolymeric 
term since this energetic contribution is a function only of the number of 
contacts, irrespective of whether they were native or not.
The extra gap energy conveniently defined with respect to fully 
collapsed states in (\ref{eq:fqt}) is an energetic contribution added to 
each native bond formed, independent of $\zbar$, up to the limit $Q\zn=\zbar$
set by (\ref{eq:etaH}) , where the gap term in (\ref{eq:fqt}) becomes
simply $\zbar\delen$.

These substitutions in (\ref{eq:fqt}) describe a free energy surface as a
function of the reaction coordinates $Q$ and $\zbar$ - essentially the 
native contacts and the total contacts:
\bgeq
\frac{F}{N}\left(T,Q,\zbar |E_n\right) =
	- \zbar \left| \ebar \right| - Q \zn \left| \delen \right|
	- T s\left( Q, \zbar \right) - 
	\frac{\zn \etaHqz \varepsilon^2}{2 T} \left(1-Q^2\right) \; ,
\label{eq:fqz}
\endeq
where $s\left( Q, \zbar \right) = 
s\left( Q, \etaH^{tot}\left( Q,\zbar \right) \right)$. 
The first term is an equilibrium bias towards states that simply have more 
contacts and depends only on $\zbar$, whereas the second term is a bias
towards states with greater nativeness and is depends only on $Q$,
although the maximum value of this bias does depend on $\zbar$ as 
explained above.  The entropic term biases the free energy minimum
towards both small vlaues of $Q$ and $\zbar$ where the entropy is largest.
The energetic parameter $T$ determining the weight of this term is held fixed 
at a value $\tf$ described below, and the other energetic parameters 
($\ebar$, $\delen$, and $\varepsilon$) are 
adjusted so as to give the free energy a double well structure with
folded and unfolded minima of equal depth.  The free energy bias due to
landscape roughness is largest when there are many 
non-native contacts ($\zbar$ is large and $Q$ is small) which means that
the protein can find itself in non-native low energy states due to the 
randomness of those non-native interactions.

\subsection{Comparison with a Simulation}

The $27$-mer lattice model protein has been simulated for polymer
sequences designed to show minimal frustration.~\cite{SO,SOpriv} 
The system we are interested in 
is modelled by a contact hamiltonian as in (\ref{eq:ham}),
but now the beads representing the monomers are of $3$ different kinds 
with respect to their energies of interaction.
If like monomers are in contact, they have an energy 
$\varepsilon_{ij} = -3$, otherwise $\varepsilon_{ij} = -1$,
where the interaction energies are in 
an arbitrary scale of units of order $k_{\mbox{\tiny B}} T$. 
This specific sequence is modelled to have 
a fully collapsed ``native'' state with a specific set of 
$28$ contacts and a
ground state energy of $-3 \times 28$.

In the thermodynamic limit, the discrete 
interaction energies used in the 
simulation give a gaussian distribution for the total energy of the system
by the central limit theorem, whose mean and width naturally depends on the 
fraction of native contacts.

If we call $\Zbar$ the total number of contacts of any kind,
the energy at $Q$ and $\Zbar$ is determined simply by the energies of 
these native and non-native contacts above, while the entropy at high 
temperatures is the log of the number of states satisfying the 
constraints of $\Zbar$ total contacts and $\mu$ native contacts.
However, the temperature range where folding occurs is well below the temperature
of homopolymeric collapse, and so the polymer can be considered to be
largely collapsed. This can be seen either by direct computation or
by computing the entropy~\cite{SOpriv}, defined 
through
\bgeqarray
S(Q,\zbar,T) &=& -\sum_i p_i \log p_i \nonumber \\
	&=& -\sum_i \left(\frac{\mbox{e}^{-E_i/T}}{Z_p}\right)
	\log \left(\frac{\mbox{e}^{-E_i/T}}{Z_p}\right)
\endeqarray
where $Z_p$ is the (partial) partition function, 
the sum being over all of the 
states consistent with the constraints 
characterized by $\mu$ and $\Zbar$ above. 

We can now easily obtain the free energy as $F=E-TS$, shown in
Fig.~\ref{fig:data}A for the $27$-mer as a surface plot 
vs. the total number of contacts 
per monomer $\zbar = \Zbar/N$, and $Q = (\mbox{total number of native 
bonds})/28$, the number of 
native contacts over the total
number of possible contacts~\cite{owls}. The largest value of $Q$ for a given 
$\Zbar$ is $\Zbar/28$, because there cannot be more native contacts than 
there are total contacts, hence the allowable region is the upper left
of the surface plot.
The surface plot in fig.~\ref{fig:data}A has a double minimum structure at a 
specific temperature $\tf=1.51$ on the energy scale where 
$\varepsilon_{ij} = \{ -3,-1\}$ described above.  
The free energy barrier of $\approx 2 \kboltz \tf$ 
is small compared with the entropic barrier of the system
( $\sim 14 \, \kboltz \tf$). The transition ensemble 
at reaction coordinates $(\qst,\zst) \cong (0.54,0.88)$, consists of about
$\exp N s(\qst,\zst) \cong 2,000$ thermally occupied states and  
$\sim 10^5$ configurational states.

There are $4$ energetic parameters in the free energy theory ($\varepsilon,
\ebar, \delen$, and $\kboltz \tf$), and 
$3$ parameters in the simulation ($\varepsilon (\mbox{like units}),
\varepsilon (\mbox{unlike units})$, and $\kboltz \tf$), 
plus the roughness parameter, which is 
implicitly evaluated through the diversity of energies
consistent with overlap $Q$.
It is worth noting that the minimal frustration in the lattice simulation is 
implicit in the sequence design, in that the ground state is topologically
consistent with all the pair interactions between like monomers.
However, the gap energy in the simulation is functionally different 
than the theoretical model in that 
contacts between like monomers are always favored whether native or 
not, and in the theory only true native contacts have explicit contributions to
the energy gap.  This means that denser states are weighted more strongly 
in the simulation than in the theory, and thus we may expect our
homopolymer attraction parameter $\ebar$ to be somewhat larger than 
the simulational average ($\ebar \approx 2$) for the same $\tf$.

We do not undertake here a comparison of simulations at all parameter
values with theory.  Rather, we compare simulations and theory only for the 
$27$-mer, with parameters chosen to be protein-like acording to the corresponding
states principle analysis of Onuchic et. al.~\cite{owls}.
The scheme for comparison between the simulations and theory for the $27$-mer
is to hold $\tf$ fixed at the simulational value of $1.5$, 
and then determine the remaining three energetic parameters 
($\varepsilon^2 , \delen , \ebar $) by appropriate constraints.

One systematic method of finding protein-like energy parameters is to
assume the folded state is the native state, and solve three linear equations
in the energy parameters determined by the condition of folding equilibrium
\bgeqarray	
	F\left( Q_g , \zbar_g \right) &=& F \left( Q_{fold} , \zbar_{fold} \right)
	\nonumber \\  &=& F \left( 1 , \zn \right)  \nonumber  \; ,
\endeqarray
and the conditions that the molten globule at ($Q_g,\zbar_g$)
is a free energy minimum (or saddle point)
$$
\left. \frac{\partial F}{\partial Q}\right|_{(Q_g,\zbar_g)} = 0 \;\;\;\;\;\;
\mbox{and} \;\;\;\;\;\;
\left. \frac{\partial F}{\partial \zbar}\right|_{(Q_g,\zbar_g)} = 0 \; .
$$
However, for similar reasons as in the density un-coupled formulation of the 
free energy, the folded minimum is not equivalent to the native state,
and this assumption leads to pathologies when the previous treatment
is implemented.

The introduction of a non-uniform density in the protein leads to an ensemble
of folded states consisting of a
dense core (with $\eta = 1$), and an expanded
halo (of dilute densities $\sim 0.05$), as determined by eq. (\ref{eq:etaH})
at the reaction coordinates of the folded state ensemble.
Folded states with this structure of a core containing nearly all the 
contacts, and dangling loops or ends, are not entirely inconsistent 
with what is known about real folded proteins, which consist of 
regions of sequence with 
well-defined spatial structure (e.g. the tertiary arrangement of helical 
segments) along with regions of somewhat greater entropy density with not as
well-defined structure (e.g. the ``turns'' in a helical protein). 
However, one should still keep in mind the previous comment regarding 
the simplifying assumption of a non-interacting halo in the theory of 
collapse.

Viewing the problem from a somewhat different angle, we 
can seek energetic parameters comparable to the lattice simulation values
which give a double well free energy surface in the coordinates $Q$ and $\zbar$,
 with a barrier 
position and height consistent with simulations and experiments.

The result of this is shown in figure \ref{fig:data}B, 
which shows the free energy surface at $\tf$ obtained 
from the parameters $\Evect = ( 0.9 , -2.8 , -1.6 , 1.51 )$.
The gap to roughness ratio for this minimal model is 
$|\delen | / \varepsilon \cong  1.8$ (satisfying the conditions for global
foldability).
The system has a double well structure with a 
weakly first-order transition between a collapsed
globule, at $(\qmg,\zmg) \cong (0.07 , 0.97)$, 
and a somewhat expanded ($ \approx 3$ or $4$ less contacts for the $27$-mer) 
core-halo like folded state at 
$(\qf , \zf ) \cong (0.80 , 0.83)$.  
For these energetic parameters, the folded state is more energetically
favored 
with $E_f - E_{mg} \cong -3.6 \kboltz T$ and thus less entropic 
($\tf (S_f - S_{mg}) \cong -3.6 \kboltz T$).
For the folded states, the density of the polymer is inhomogeneous,
with a core containing about $23$ monomers at density $\etaC = 1$ and 
a halo of about $4$ monomers at density $\etaH$ effectively zero.
A more exact theory would impose the constraint of chain connectivity on
the halo, which would significanty increase its density, and decrease the 
expansion effect seen here ($\Delta \zbar \cong 0.14$).  This expansion
effect is naturally reduced as the average homopolymer attraction energy
becomes larger, and is also reduced for rougher landscapes, where 
non-native contacts with large variance of interaction energy can
contribute to deeper minima. In the folded enemble,
all the entropy is essentially in the dilute, expanded halo, 
and all the energy in the dense core.

The core residues in the transition state ensemble
at $(\qst,\zst) \cong (0.5 , 0.8 )$ contain approximately
$N \zn \qst / \znq \cong 16$ monomers. Due to its position, 
the transition state ensemble 
has almost twice the thermal entropy as in the simulations, 
so that it consists
of $\sim  10^7 $ thermally occupied states and $\sim 10^8$ configurational states.
The free energy barrier at this position is about $\Delta F \cong 6 \kboltz T$ 
of which the energetic 
and entropic contributions, from equations (\ref{eq:et}) and (\ref{eq:sqt})
respectively,  are $\cong 1 \kboltz T$ and $ 5 \kboltz T$ respectively,
the entropy loss to condense the critical core being the more dominant
factor here.
The barrier height is naturally reduced for smaller homopolymeric biases, 
and also for rougher landscapes, where traversing the rugged landscape to
find the folded state becomes more of a second order, less collective process.

The core-halo expansion effect is exaggerated for
the parameters used in simulating a lattice model $27$-mer.
One explanation is that nativeness instills collapse not just in the subunits that have
native contacts, but also in the surrounding polymer medium
because of topological constraints not considered
in the simple non-interacting core-halo model.  We should also bear in 
mind that at high $Q$ for the $27$-mer there are significant lattice effects
in the simulation.  The non-self averaging behavior here cannot be 
predicted by the simple polymer model.

\subsection{Explicit $3$-body effects}

It is interesting to investigate the effects of 
explicit many-body cooperativity on the folding funnel by 
introducing a $3$-body interaction in addition to
the pair interactions already present.
Models with such partially explicit cooperativity mimic the idea
that only formed secondary structure units can couple, and have been
introduced in lattice models by Kolinski et al.~\cite{SO}.
$3$-body interactions enter into the energetic contributions of
(\ref{eq:fqz}) as an additional $Q^2$ term in the bias and roughness,
and $\zbar^2$ term in the homopolymer attraction, so that those terms 
in the free energy become
\bgeqarray
	- \left[\left(1-\alpha\right)\zbar+\alpha \zbar^2 \right] \left| \ebar \right| 
	&-& \left[\left(1-\alpha\right)Q+\alpha Q^2\right] \zn \left| \delen \right| 
	\nonumber \\
	\;\;\;\;\;\;\;\;&-& \frac{\zn \etaHqz \varepsilon^2}{2 T} \left\{1-
	\left[\left(1-\alpha\right)Q+\alpha Q^2\right]^2\right\} \; . \nonumber
\endeqarray
We can obtain the parameters characterizing the barrier as a function of the 
three-body coefficient $\alpha$.
The collectivity induced by $3$-body interactions makes 
the energetic funnel steeper and narrower, 
and the gap bias is then effective only for higher $Q$.
This means that to maintain equilibrium between the globule and folded states
the landscape must either be less rough or more strongly biased.
We choose to increase the stability gap $|\delen |$ 
at constant roughness $\varepsilon$ 
and temperature $\tf$.  
The magnitude of the gap energy is a roughly linearly increasing function of 
the coefficient of the three body term $\alpha$, rising from $1.6$ at 
$\alpha=0$ to $\cong 2.1$ at $\alpha = 1$, in units where $\tf =1.51$.
With this correction included, we find the barrier
position $\qst\left(\alpha\right)$ to be only weakly dependent on $\alpha$
(although as described above, 
$\qst$ is not independent of $m$, the order of the $m$-body interactions),
and the position of the folded state $\qf$ to weakly increase (figure \ref{fig:q3body}A).
As expected, the transition becomes more first order-like with increasing $\alpha$
(the free energy barrier increases), but there is  
a non-trivial dependence of the entropic and energetic contributions to the
barrier (see Fig. \ref{fig:q3body}B).

\subsection{Dependence of the barrier on sequence length}

It is simple in our theory to vary the polymer sequence length. One recalculates
$s(Q)$ at constant density \cite{pww} and inserts this, along 
with eq. (\ref{eq:etaH}) at the larger value of $N$, into the free energy
(\ref{eq:fqz}).  Then one must rescale the temperature since in our model
larger proteins fold at higher temperatures, i.e. By equation (\ref{eq:tf})
the folding temperature should scale with $N$ somewhat greater than as $\zn$
(see figure \ref{fig:tNplot}A). Figure \ref{fig:tNplot}B shows that the
resulting free energy has a barrier
whose position is a mildly decreasing function of $N$. 
An explanation for this is that in larger polymers, entropy loss due to 
topological constraints is more dramatic in $Q$ because a smaller fraction 
of total native contacts is necessary to constrain the polymer.
That is, as $N$ increases, the number of bonds per monomer  
in the fully constrained state ($Q = 1$) approaches the bulk 
limit of $2$ (see eq. [\ref{eq:ZofN}]), while only one bond is needed to 
constrain a monomer. So this pushes the position $\qst$ of the barrier 
in, roughly as $1/\zn$.
Plotted along 
with the theoretical curve are three experimental measurements of the 
barrier position. The square represents the measurement for
$\lambda$-repressor~\cite{oas}, a $\sim 70$ residue protein with largely helical 
structure.  The triangle represents Chymotrypsin 
Inhibitor $2$ (CI2)~\cite{fersht},
a $64$ residue protein with both $\alpha$-helices and $\beta$-sheets.
The correspoding states analysis~\cite{owls} shows that the formation of 
secondary structure within these proteins makes them entropically 
analogous to the lattice $27$-mer.
Also plotted in the figure (circle) is the experimental barrier measurement for 
Cytochrome C~\cite{gray}, a $104$ residue helical protein which is 
entropically similar to the $64$-mer lattice model.
The simple proposed model is in reasonable accord with the general 
quantitative trend in the position of the 
transition state ensemble that is observed experimentally.
In appendix A, we show for thoroughness that experimental plots of folding
rate vs. equilibrium constant 
are indeed a measure of the position of  the transition state ensemble.

Figure \ref{fig:tNplot}C shows the general increasing trend of the barrier 
height as $N$ increases, as well as the entropic and energetic contributions 
to the barrier. The increasing entropy and decreasing energy of the barrier
indicate a more significant expansion with $N$ 
at the transition state coordinates $(\qst,\zst)$. In this model larger proteins 
expand more to rearrange the backbone to the folded three-dimensional 
structure.

\subsection{Dependence of the barrier on the stability gap, at
fixed temperature and roughness.}

As the stability gap is increased at fixed temperature, 
folding approaches a downhill process, with the folded ensemble becoming
the global equilibrium state (see Fig.~\ref{fig:FqstVsEn}A).
We can see from figure \ref{fig:FqstVsEn}A that the barrier position 
and height are decreasing functions of stability gap, with true
downhill folding (zero barrier) occuring when $\delen/\varepsilon \cong
2.4$ or $\delen/\tf s_o \cong 1.6$ for the $27$-mer 
(see Fig.s~\ref{fig:FqstVsEn}B and \ref{fig:FqstVsEn}C).
At $\tf$, $\delen/\tf s_o \cong 1.2$. Thus, achieving downhill folding 
requires a considerable change of stability - an estimate for a 
$60$-mer protein would be an excess stability of $\approx 8 \kboltz \tf$.

We can apply the equations of Appendix A to changes of the transition
state free energy by modifying stability.
Fig.~\ref{fig:lnlnK}A
shows a plot of the log of a normalized folding rate 
$\ln k$ vs. the log of the 
unfolding equilibrium constant $\ln K_{eq}$, whose slope is a measure 
of the barrier position $\qst$.  The increasing magnitude of slope 
with increasing $\ln K_{eq}$ means that the barrier position is shifting
towards the native state as the gap decreases.  Figure~\ref{fig:lnlnK}B
shows the actual position of the barrier, along with $\qst$ as derived 
from the slope of figure \ref{fig:lnlnK}A from eq.~(\ref{eq:slope2}).

\subsection{Denaturation with increasing temperature.}

The probability $P_u$ for the 
protein to be in the unfolded globule state at temperature $T$ is
$$
P_u = \left[1 + \mbox{e}^{-\frac{1}{T}\left(F_f-F_u\right)} \right]^{-1}
$$
where $F_u$ and $F_f$ are the free energies at temperature $T$ of the 
unfolded and folded minima (note that at $\tf$, $ F_f=F_u$ and $P_u=1/2$).
This is used to obtain denaturation curves.
For illustration, we make the simplifying assumptions that both the
folded and globule states are collapsed, making $P_u$ independent of
$\ebar$, and that the folded and globule states occur approximately
at $\qf = 1$ and $\qmg = 0$.  As the temperature is lowered, 
the molten globule freezes into a low energy configuration at $T_g
= \varepsilon \sqrt{\zn \left(1-Q_{mg}^2\right)/
\left(  2 s\left(Q_{mg}\right)\right) }\cong \varepsilon \sqrt{\zn /\left(
2 s_o \right)}$ (see Fig.~\ref{fig:tgQZ}), and the 
expression for $P_u$ becomes one of equilibrium between two temperature 
independent states with the corresponding ``Shottky'' form of the energy
and specific heat:
\bgeqarray
P_u &=& \left[1 + \exp\left(-N s_o\right) 
	\exp N\zn\left(\frac{\left|\delen\right|}
	{T} - \frac{\varepsilon^2}{2 T^2} \right) \right]^{-1}
	\;\;\;\;\;\;  T_g < T	\nonumber \\
	&=& \left[1 + \exp \frac{N\zn}{T}\left(\left|\delen\right| - 
	\varepsilon \sqrt{\frac{2 s_o}{\zn}}\right) \right]^{-1}  
	\;\;\;\;\;\;\;
	\;\;\;\;\;\;\;\;\;\;\;\;\;\;\;  T<T_g \; .
\label{eq:pd}
\endeqarray
The condition that $\tf/\tg \geq 1$ gives eq. (\ref{eq:foldability}).
Using the glass temperature of the globule state, this is equivalent to
$$
T_g \geq \frac{\varepsilon^2}{\delen} \;\; ,
$$
which is the temperature where the high $T$ expression for $P_u$ has a
minimum.  Hence cold-denaturation will not be seen in the constant 
density model (as it would if there were no glass transition), 
and $P_u$ will always decrease to zero at low temperatures.

In the limit of large $T$, (\ref{eq:pd}) becomes $\approx 1/\left( 1+ \exp 
-N s_o \right) \approx 1$, indicating denaturation. 
At small $T$ (\ref{eq:pd}) tends to zero as
$\exp -\left(\mbox{const.}\times N/T\right)$.
The denaturation curve from equation (\ref{eq:pd}) 
is plotted in figure \ref{fig:pdenat2}
for two proteins of different roughness. The ratios of widths to
folding temperatures $\Delta T/\tf$ are about $0.2$ and $0.4$
for the parameters used in the figure. These are consistent 
with the simulation values.~\cite{soKin}

The expanded halo of the native states modifies 
the denaturation behavior. The increased
entropy of the folded state with a shrunken core
leads to a partially re-entrant folding transition, which is 
quite weak.  We believe this to be an artifact of the over-estimated
entropy of the halo.

\section{Conclusion}
\setcounter{equation}{0}

In this paper we have shown that if the energy of a given 
configuration of a random heteropolymer is known to be lower than 
expected for the ground state of a completely random sequence (i.e. 
the protein is minimally frustrated), then correlations in the energies
of similar configurations lead to a funnelled landscape topography.  
The interplay of entropic loss and energetic loss as the system approaches 
the native state results in a free energy surface with two-state behavior
between an unfolded globule of large entropy, and a folded ensemble of 
lower energy but non-negligible entropy. 
The weakly first-order 
transition is characterized by a
free energy barrier which functions as a ``bottle-neck''
in the folding process.

The barrier is small compared with the total 
thermal energy of the system - on the 
order of a few $\kboltz T$ for smaller proteins of sequence length about $60$
monomers.
For these small proteins the model predicts a position of the barrier $\qst$ 
about half-way between the unfolded and native states ($\qst \cong 1/2$).
For larger proteins, the barrier in $Q$ (the fraction of native contacts)
moves further away from the native state
towards the molten globule ensemble roughly as $1/\zn$, 
due essentially to the fact that the entropy decrease per contact
is indepent of $N$ initially.  Experimental
measurements of the barrier for fast folding proteins
are consistent with this predicted shift in position 
with increasing sequence length.

The folded, unfolded and transition states are not single configurations 
but ensembles of many configurations.  The transition state ensemble 
according to this theory
consists of about $10^7$ thermally accessible states for a small protein
such as $\lambda$-repressor, roughly the combinatorial number of $16$-mer core
residues in the transition state ensemble of the $27$-mer minimal model.
This observation is in harmony with the picture of 
a generally de-localized ensemble of transition state nuclei 
in contact sequence space, a subject investigated recently by various 
authors~\cite{fersht2,BB,OW}.

A simple theory of collapse was introduced to couple protein density with 
nativeness. This resulted in an expansion due to dangling residues 
rather than contraction in 
density in the process of folding. The expansion is 
overestimated due to the 
neglect of some of the effects of chain-connectivity 
in the present halo model, and a poor description of the lattice-dependent
high $Q$ protein topologies.
During folding, a dense inner native core forms, 
which grows while possibly
interchanging some native contacts with others upon completion of folding.
This core is surrounded by a halo of non-native polymer which expands 
in the folding process.  The folded free energy minimum is only about 
$80\%$ native in the model when parameters are chosen to fit simulated free energy curves for the $27$-mer.

Explicitly Cooperative interactions were 
shown to enhance the first-order nature of the 
transition through an increase in the size of the barrier, and a 
shift towards more native-like transition state 
ensembles (i.e. at higher $\qst$).
For the constant density scenario the barrier becomes almost entirely 
entropic when the order $m$ of the $m$-body interactions becomes large, and 
the transition state ensemble becomes correspondingly more native-like. 
In the energy landscape picture, as explicit cooperativity increases,
the protein folding funnel disappears, and the landscape tends towards
a golf-course topography with energetic correlations less effective and 
more short range in $Q$ space.
The correlation of stability gaps and $\tf/\tg$ ratios with kinetic 
foldability is true only for fixed $m$ much less than $N$.

A full treatment of the barrier as a function of the $3$ energetic 
parameters $(\varepsilon , \ebar ,\delen )$ plus temperature $T$ 
would involve the analysis of a multi-dimensional surface defining folding 
equilibrium in the space of these parameters.
We shall return to this issue in the future, but we have deferred it for 
now in favor of the  simpler analysis of seeking trends in the position
and height of the barrier as a function of individual parameters such as
$\delen$ and $T$.

\subsection*{Acknowledgments}
We wish to thank J. Onuchic, J. Saven, Z. Luthey-Schulten, 
and N. Socci for helpful
discussions. This work was supported NIH grant 1RO1 GM44557, and NSF grant
DMR-89-20538.

\renewcommand{\thesection}{\Alph{section}}
\setcounter{section}{0}
\section{Appendix A}
\setcounter{equation}{0}

For small, $Q$ dependent changes in the free energy (\ref{eq:fqz}), 
e.g. changes in
temperature, we can approximate the change in the free energy
at the position of the barrier $\delta F\left(\qst\right)$ as a linear 
interpolation between the free energy changes of the unfolded and folded 
minima $\delta F_u$ and $\delta F_f$, 
here estimated to be at $Q_u \cong 0$ and  $Q_f \cong 1$ respectively
\bgeq
\delta F\left(\qst\right) \cong \qst \delta F_f + \left(1-\qst\right)\delta F_u \; .
\label{eq:linF}
\endeq
Furthermore let us approximate the kinetic folding time by the 
thermodynamic folding time~\cite{OW}
$$
\tau = \overline{t\left(\qst\right)} \mbox{e}^{\left(F\left(\qst\right)-
	F_u\right)/\kboltz T}
$$
where $\qst$ is the position of the barrier and $\overline{t\left(\qst\right)}$
is the lifetime of the microstates of the transition enemble. Then the log
of the folding rate $k$ is $\propto F_u - F\left(\qst\right)$. 
The equilibrium constant for the unfolding transition $K_{eq}$ 
is the probability to be in the unfolded minimum
over the probability to be in the folded minimum, and so 
$\ln K_{eq} \propto F_f - F_u$. If we plot $\ln k$ vs. $\ln K_{eq}$,
the assumption of a linear free energy relation (\ref{eq:linF})
and a stable barrier position results in a linear dependence of rate upon
equilibrium constant, with slope
\bgeq
\frac{\delta \left[F_u - F\left(\qst\right)\right]}{\delta \left[
	F_f - F_u\right]} = \frac{\delta F_u - \delta F\left(\qst\right)}
	{\delta F_f - \delta F_u} =  - \qst
\label{eq:slope1}
\endeq
so that experimental slopes of folding rates 
vs. unfolding equilibrium constants
are indeed a measure of the position of the barrier in our theory. 

If the unfolded and folded states are not assumed to be at $Q=0$
and $Q=1$ respectively,
equations (\ref{eq:linF}) and (\ref{eq:slope1}) are modified by 
\bgeq
\delta F\left(\qst\right) \cong \left( \frac{\qst - \qu}{\qf - \qu} \right)
    \delta F_f + \left( \frac{\qf - \qst}{\qf - \qu} \right)\delta F_u \; .
\label{eq:linF2}
\endeq
and 
\bgeq
\frac{\delta \left[F_u - F\left(\qst\right)\right]}{\delta \left[
	F_f - F_u\right]} = - \left(\frac{\qst - \qu}{\qf - \qu}\right)
\label{eq:slope2}
\endeq
where $\qu$ and $\qf$ are the respective positions of the unfolded and 
folded states.  So one can obtain the barrier position from 
the slope of a plot of $\ln k$ vs. $\ln K_{eq}$, given the positions 
of the unfolded and folded states.

\newpage
\section*{Figure Captions}
\begin{figure}
\centerline{\psfig{figure=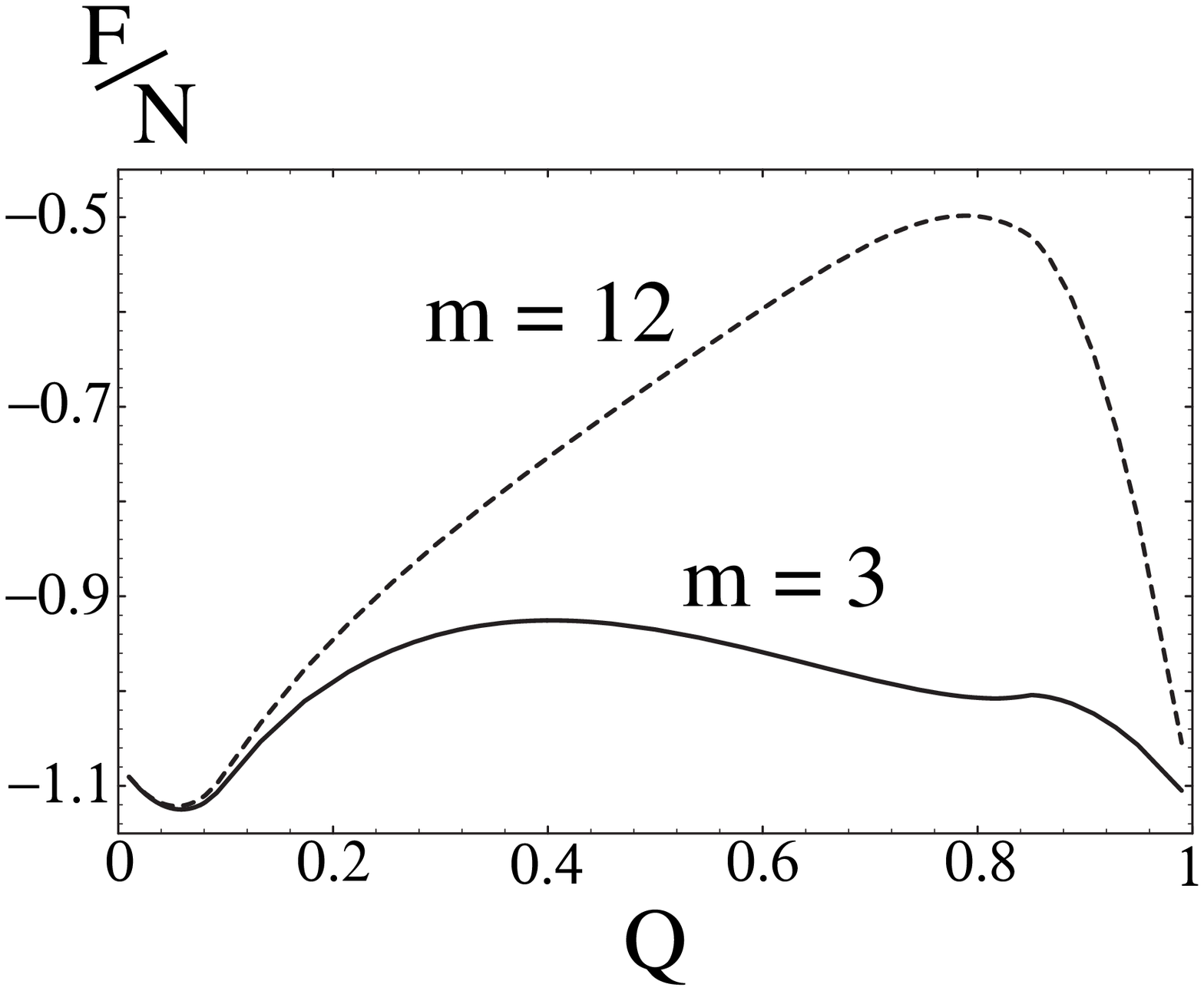,width=0.8\textwidth,angle=0}}
\centerline{\psfig{figure=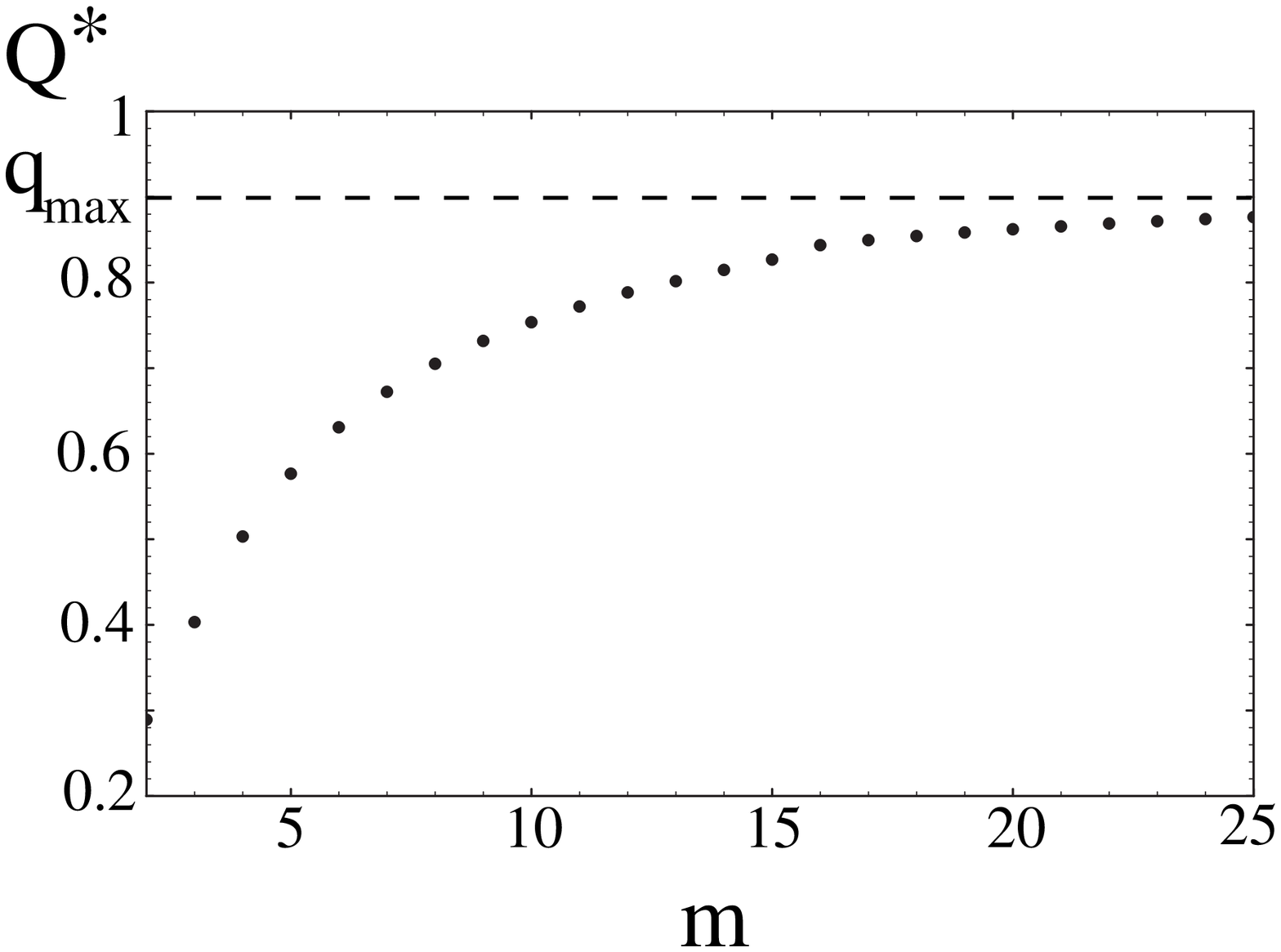,width=0.8\textwidth,angle=0}}
\centerline{\psfig{figure=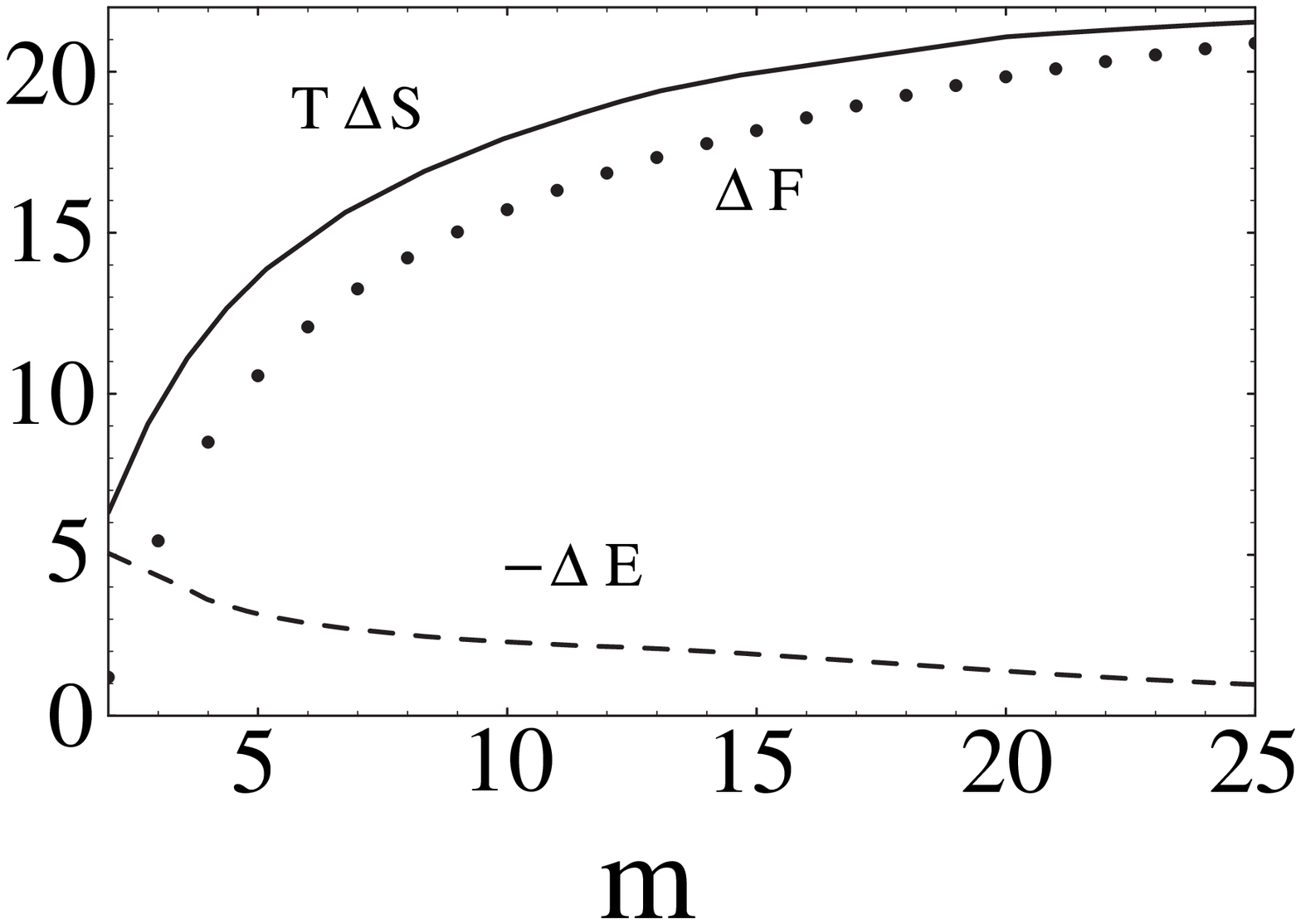,width=0.8\textwidth,angle=0}}
\caption{}
\label{fig:FvsQ3and12}
\end{figure}

\begin{figure}
\centerline{\psfig{figure=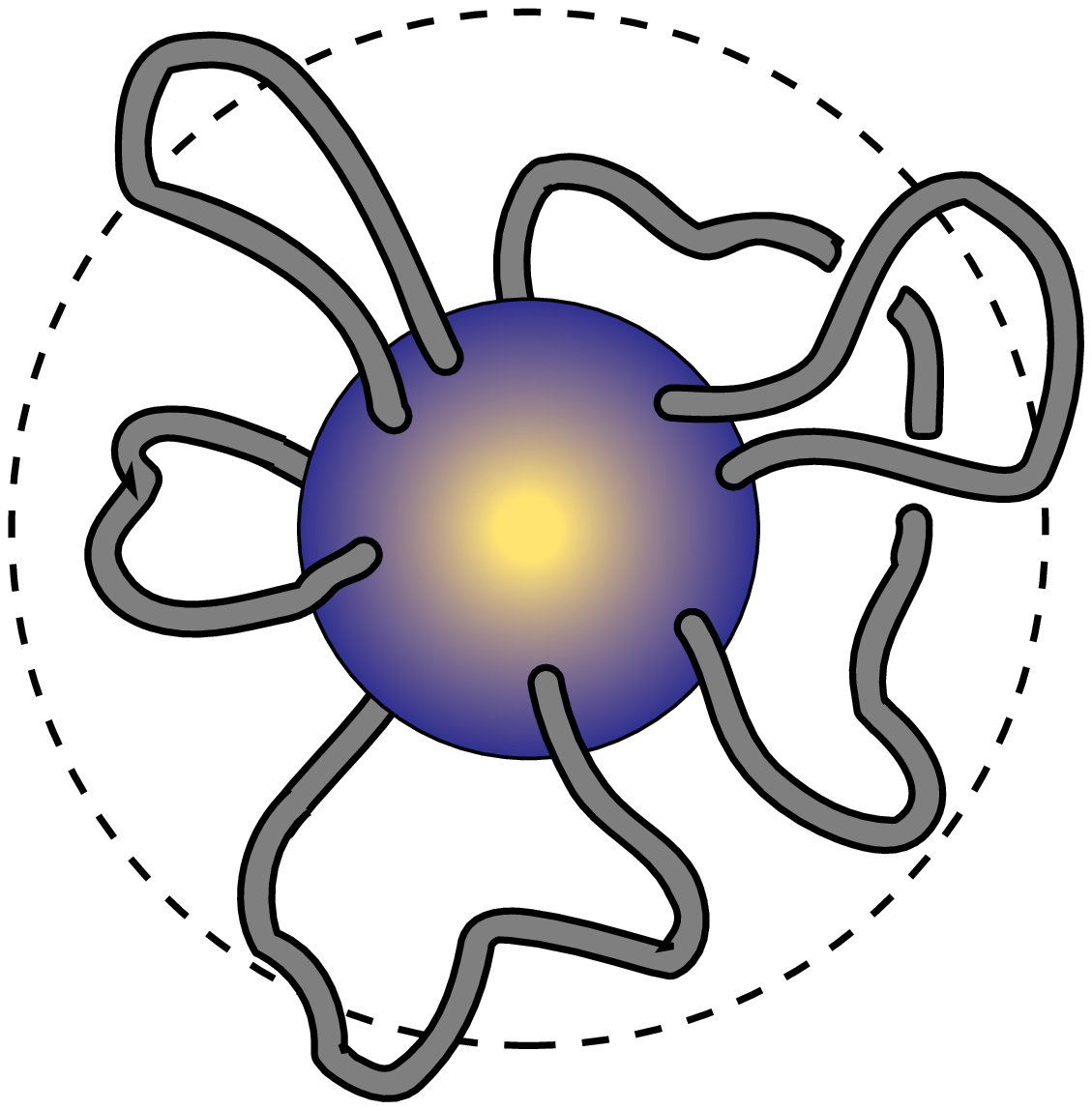,width=0.8\textwidth,angle=0}}
\caption{}
\label{fig:corehalo}
\end{figure}

\begin{figure}
\centerline{\psfig{figure=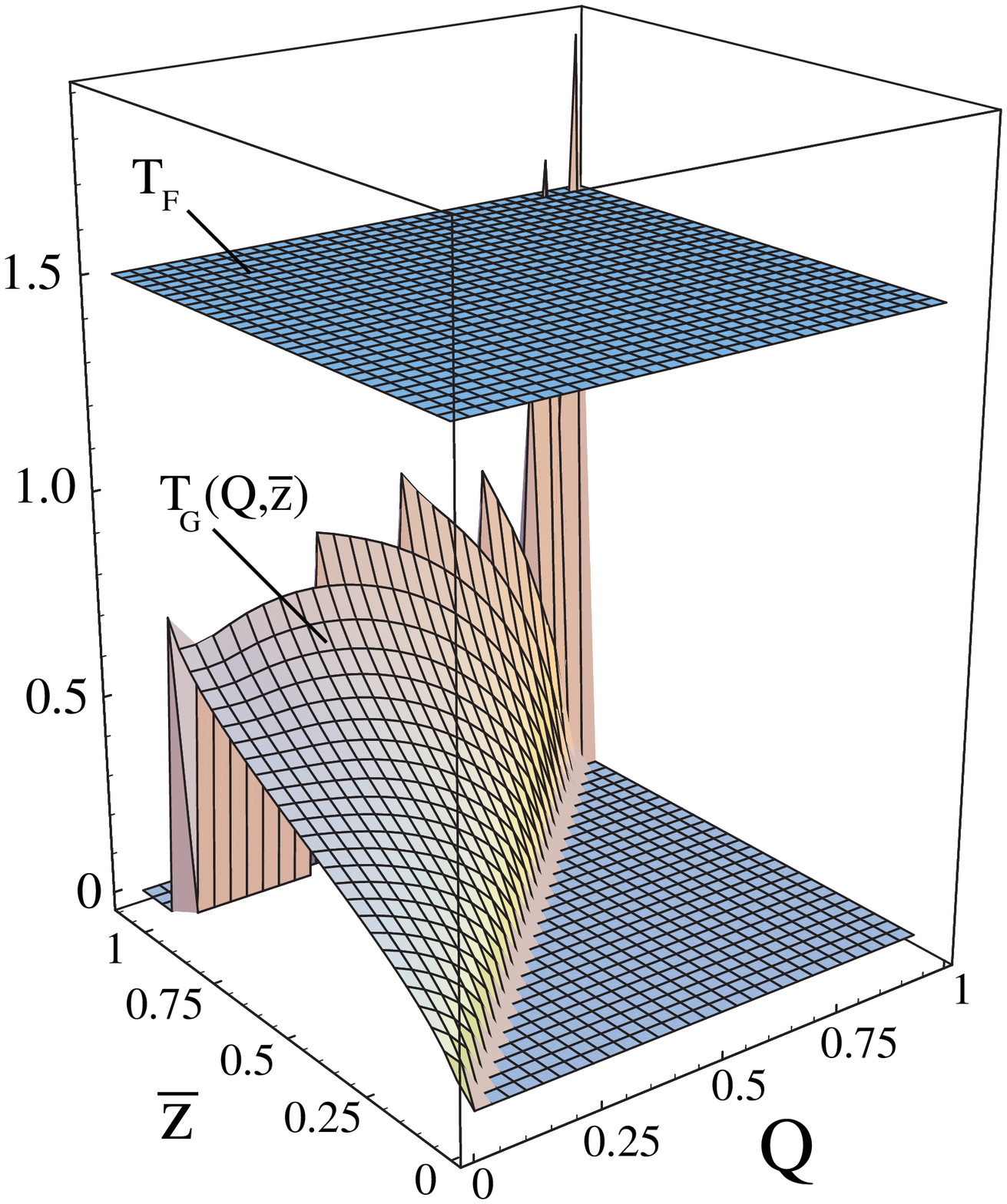,width=0.8\textwidth,angle=0}}
\caption{}
\label{fig:tgQZ}
\end{figure}

\begin{figure}
\centerline{\psfig{figure=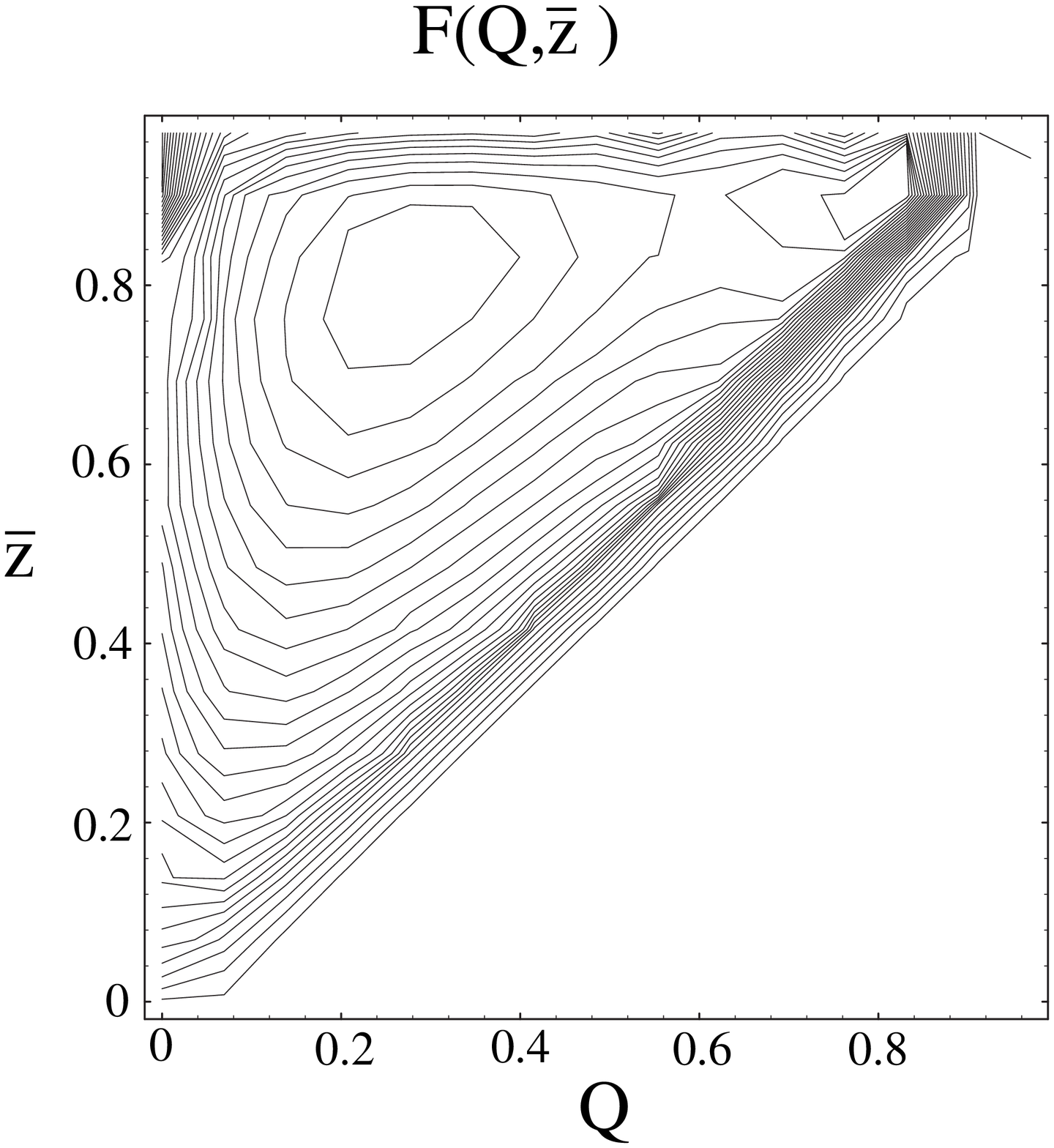,width=0.8\textwidth,angle=0}}
\centerline{\psfig{figure=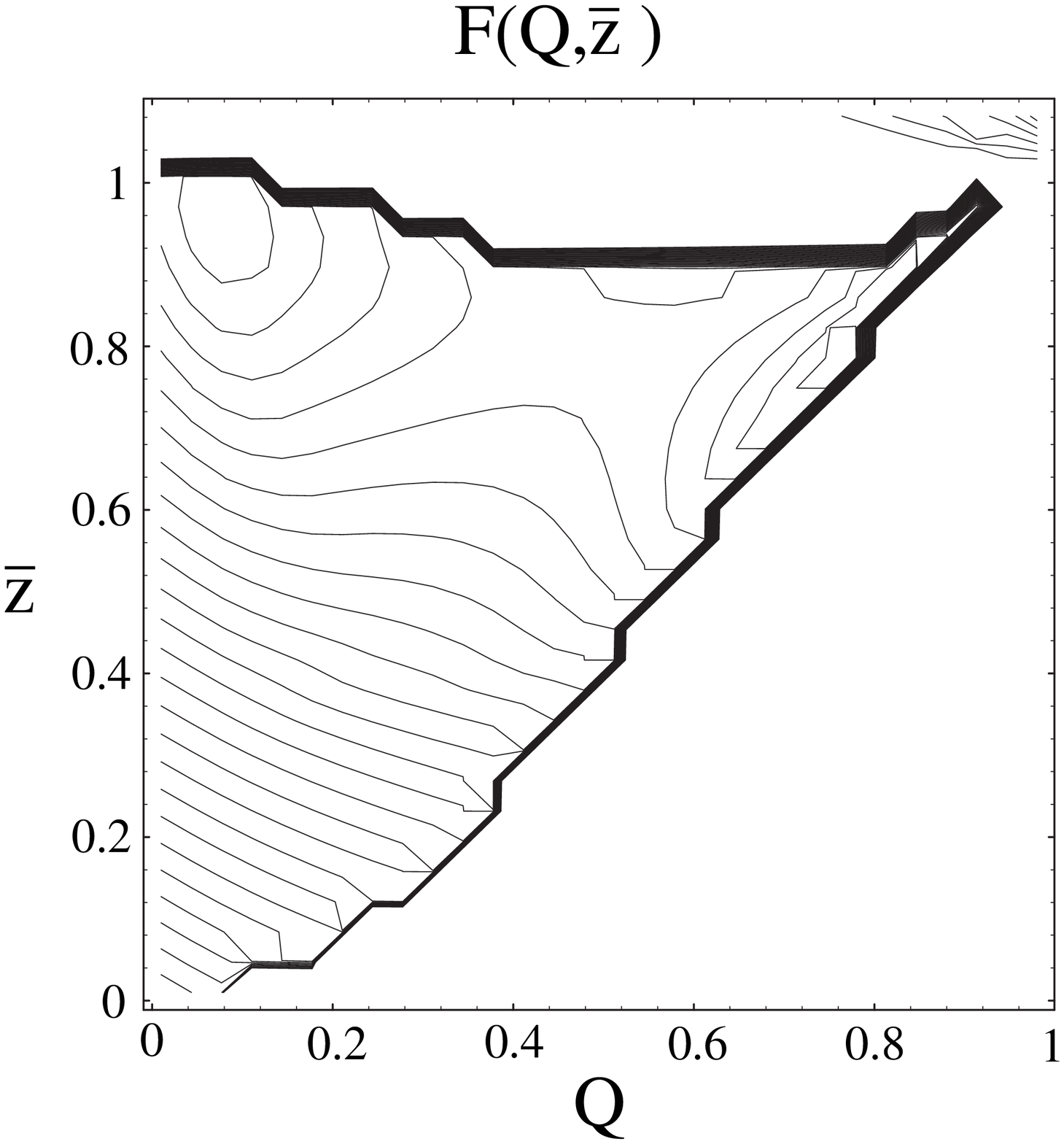,width=0.8\textwidth,angle=0}}
\caption{}
\label{fig:data}
\end{figure}

\begin{figure}
\centerline{\psfig{figure=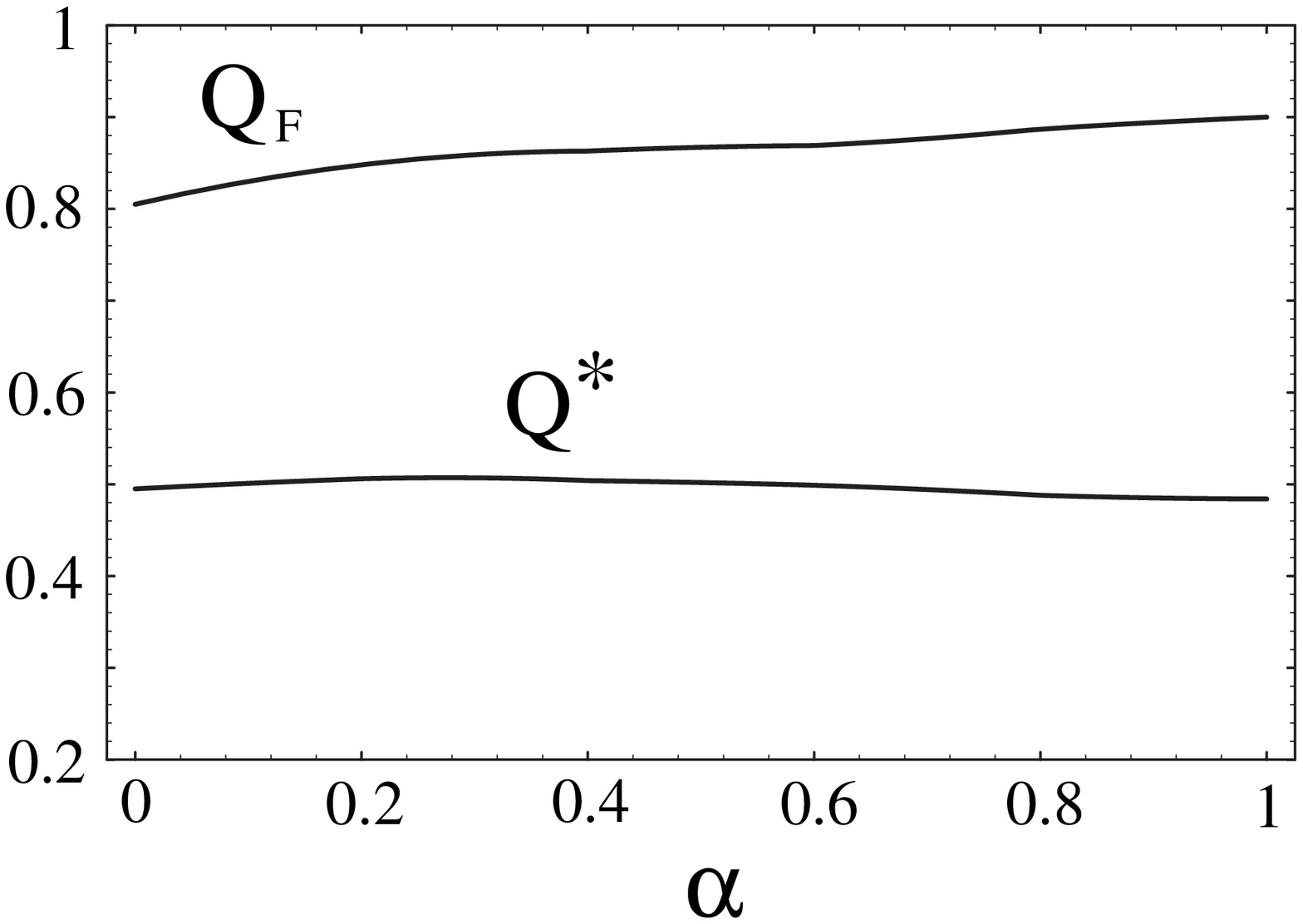,width=0.8\textwidth,angle=0}}
\centerline{\psfig{figure=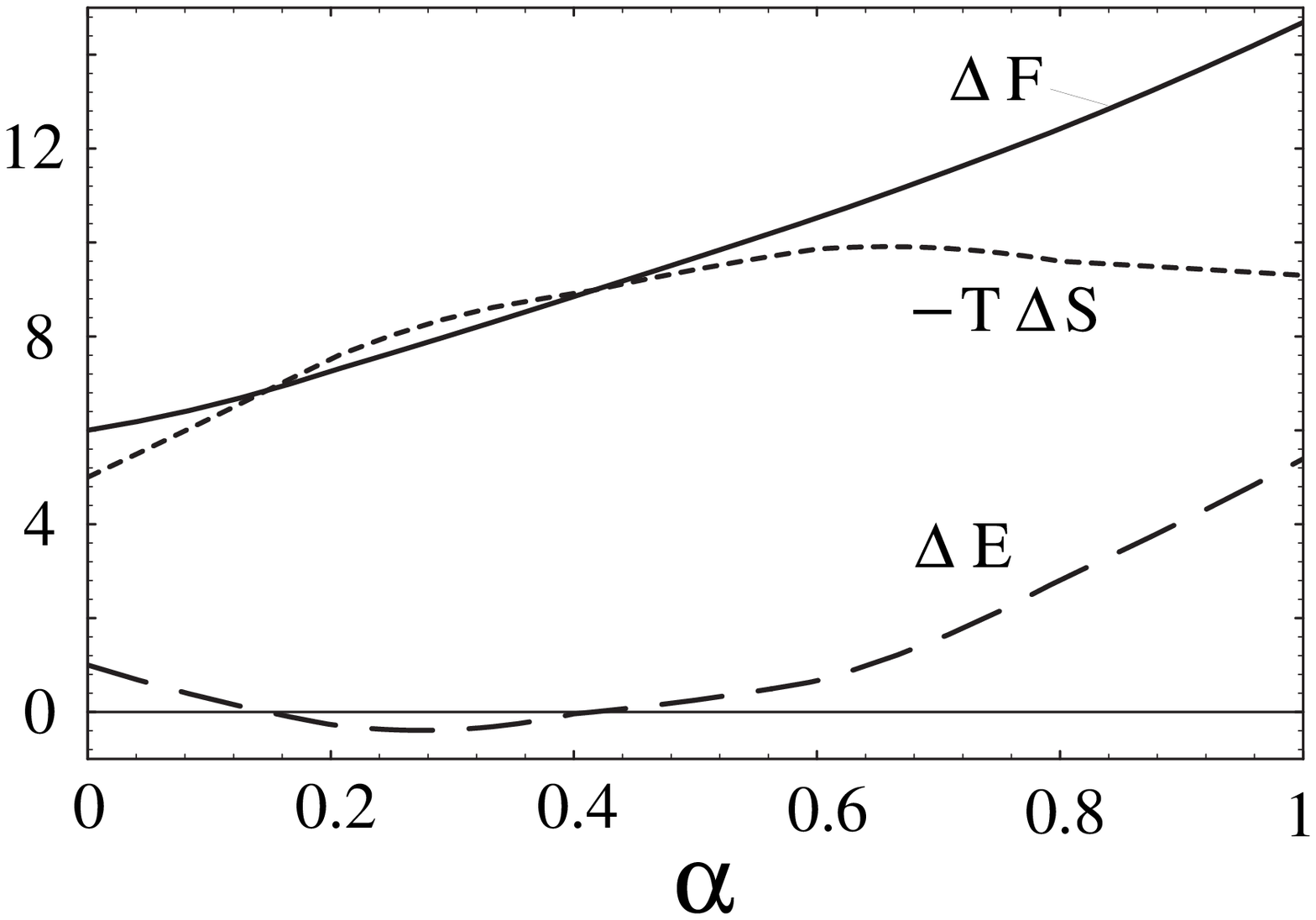,width=0.8\textwidth,angle=0}}
\caption{}
\label{fig:q3body}
\end{figure}

\begin{figure}
\centerline{\psfig{figure=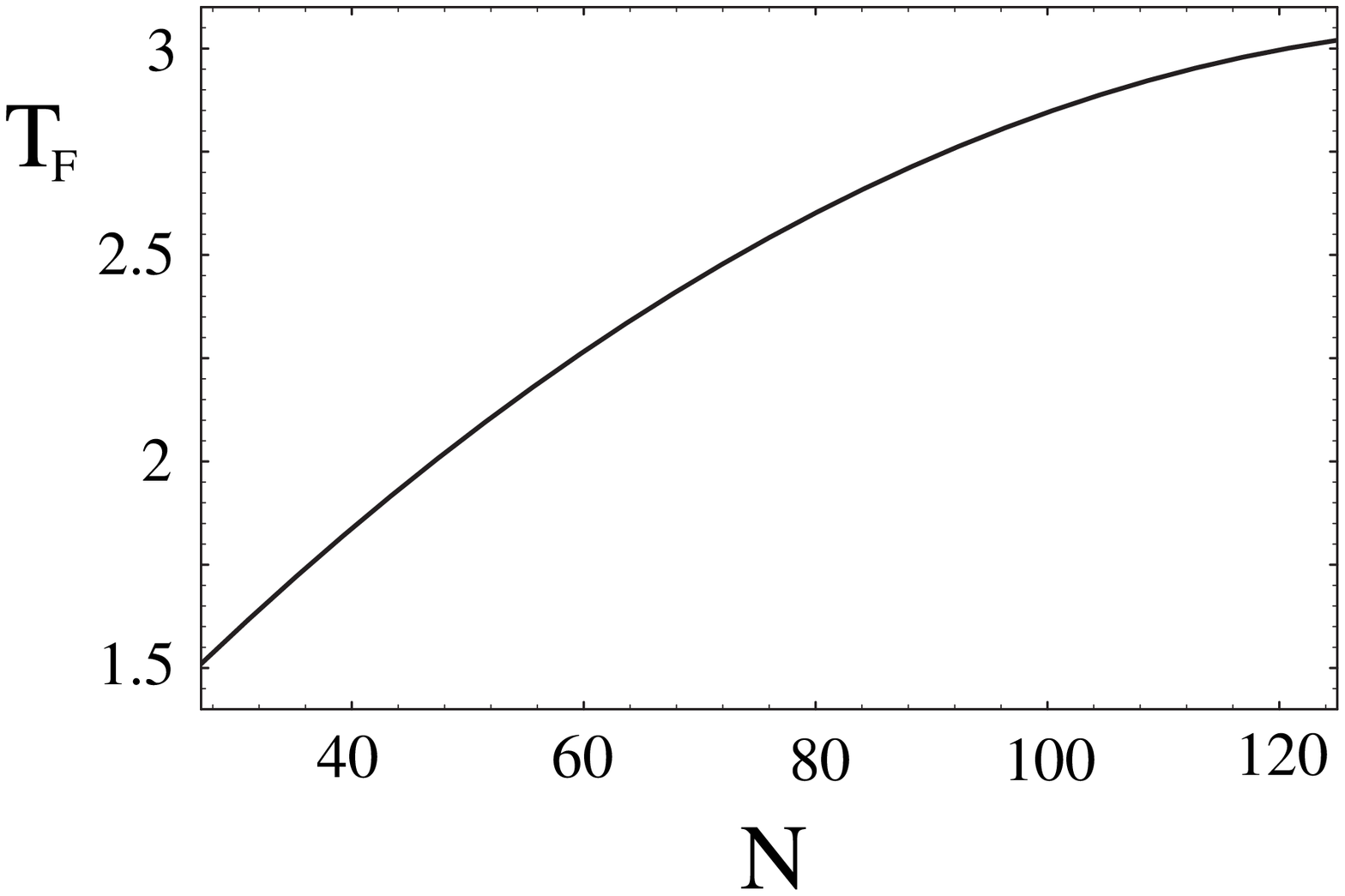,width=0.8\textwidth,angle=0}}
\centerline{\psfig{figure=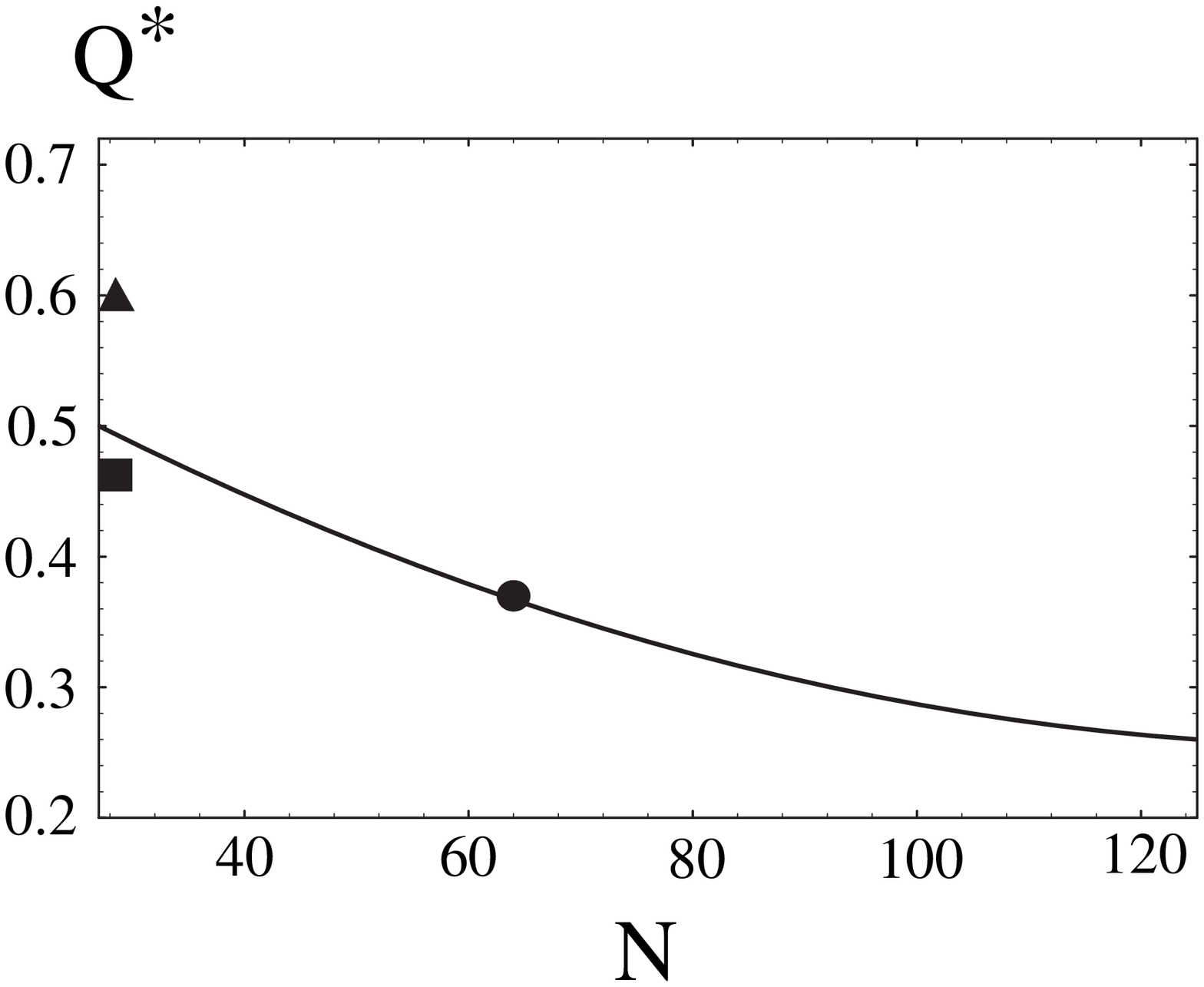,width=0.8\textwidth,angle=0}}
\centerline{\psfig{figure=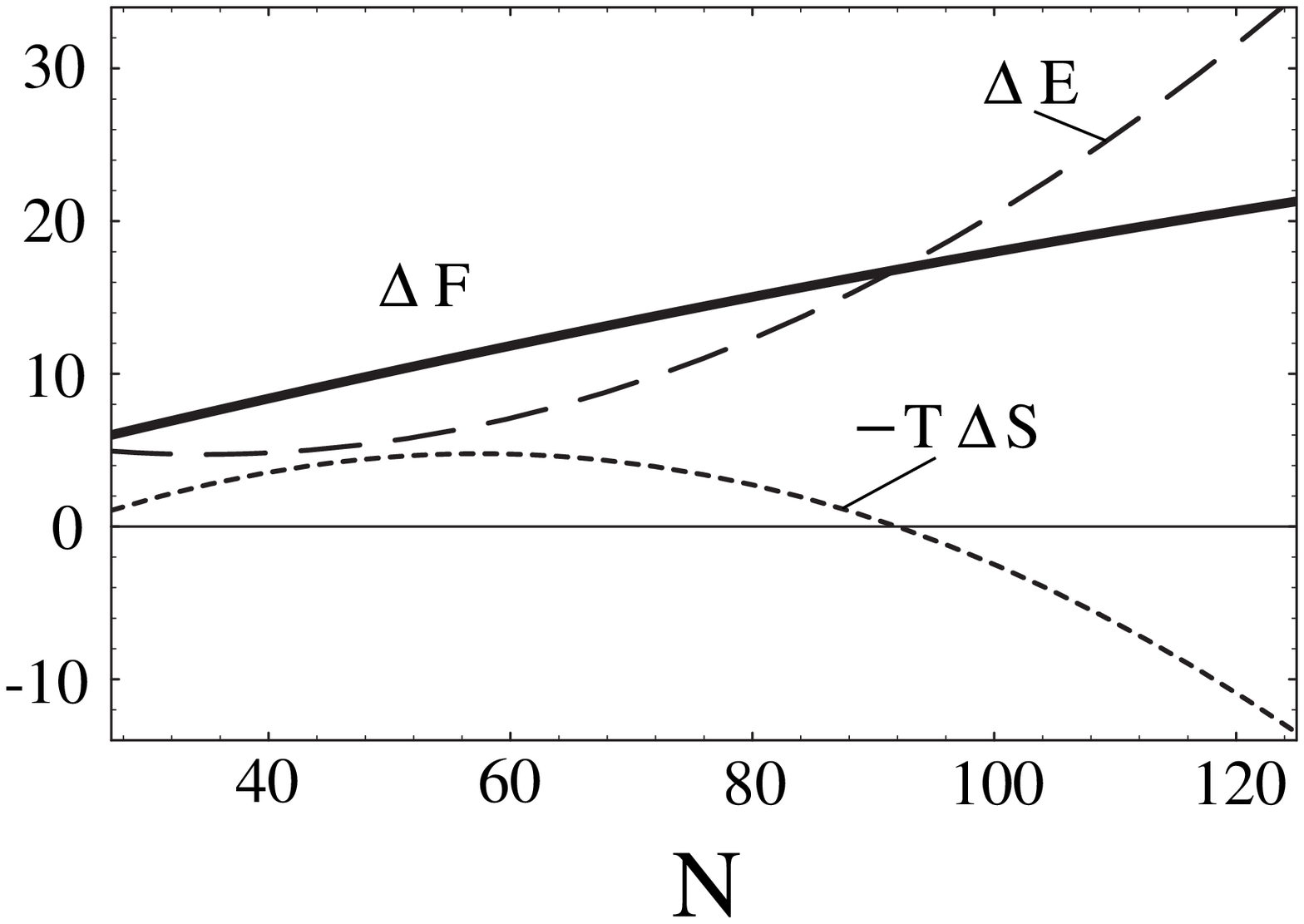,width=0.8\textwidth,angle=0}}
\caption{}
\label{fig:tNplot}
\end{figure}

\begin{figure}
\centerline{\psfig{figure=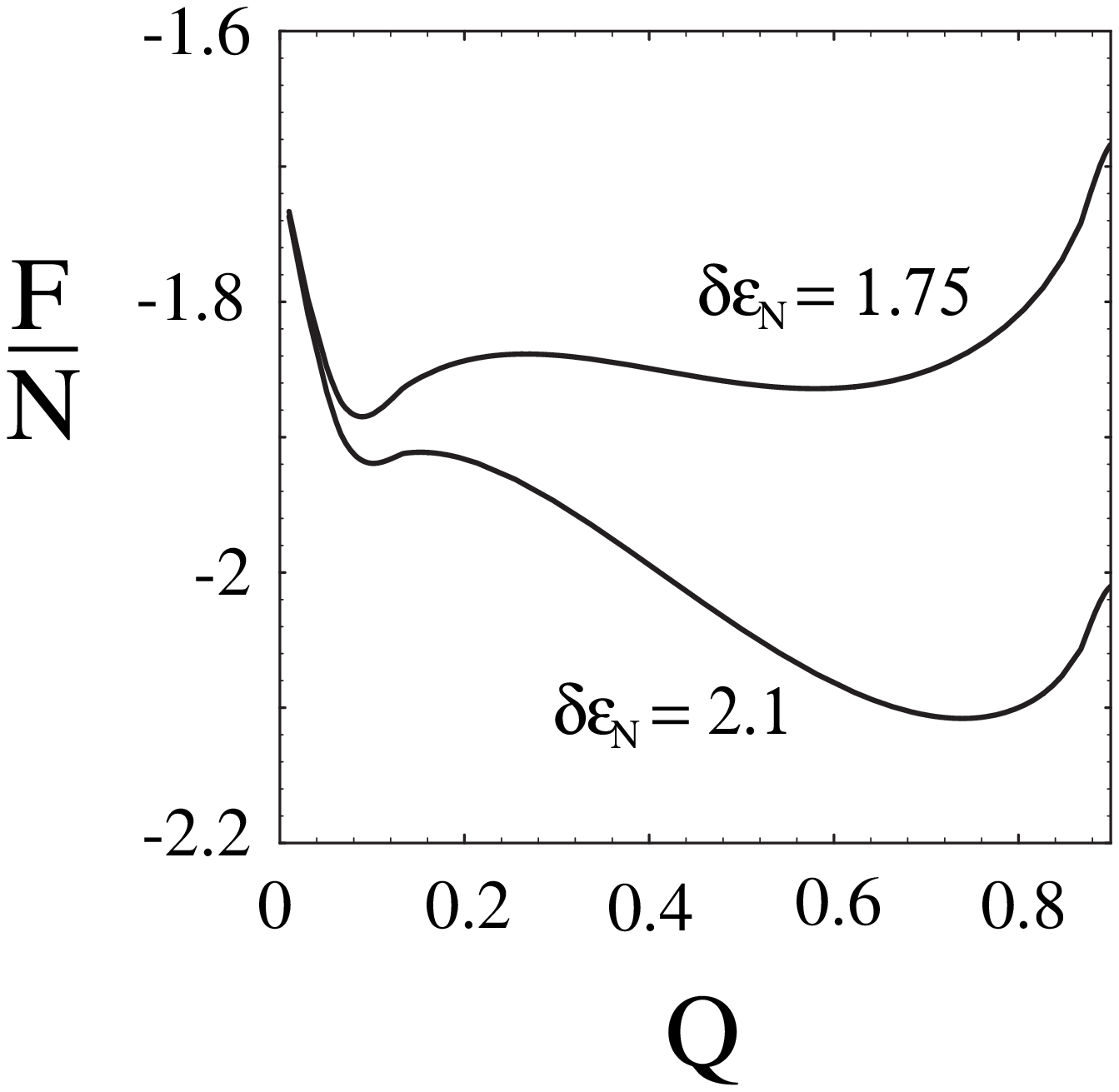,width=0.8\textwidth,angle=0}}
\centerline{\psfig{figure=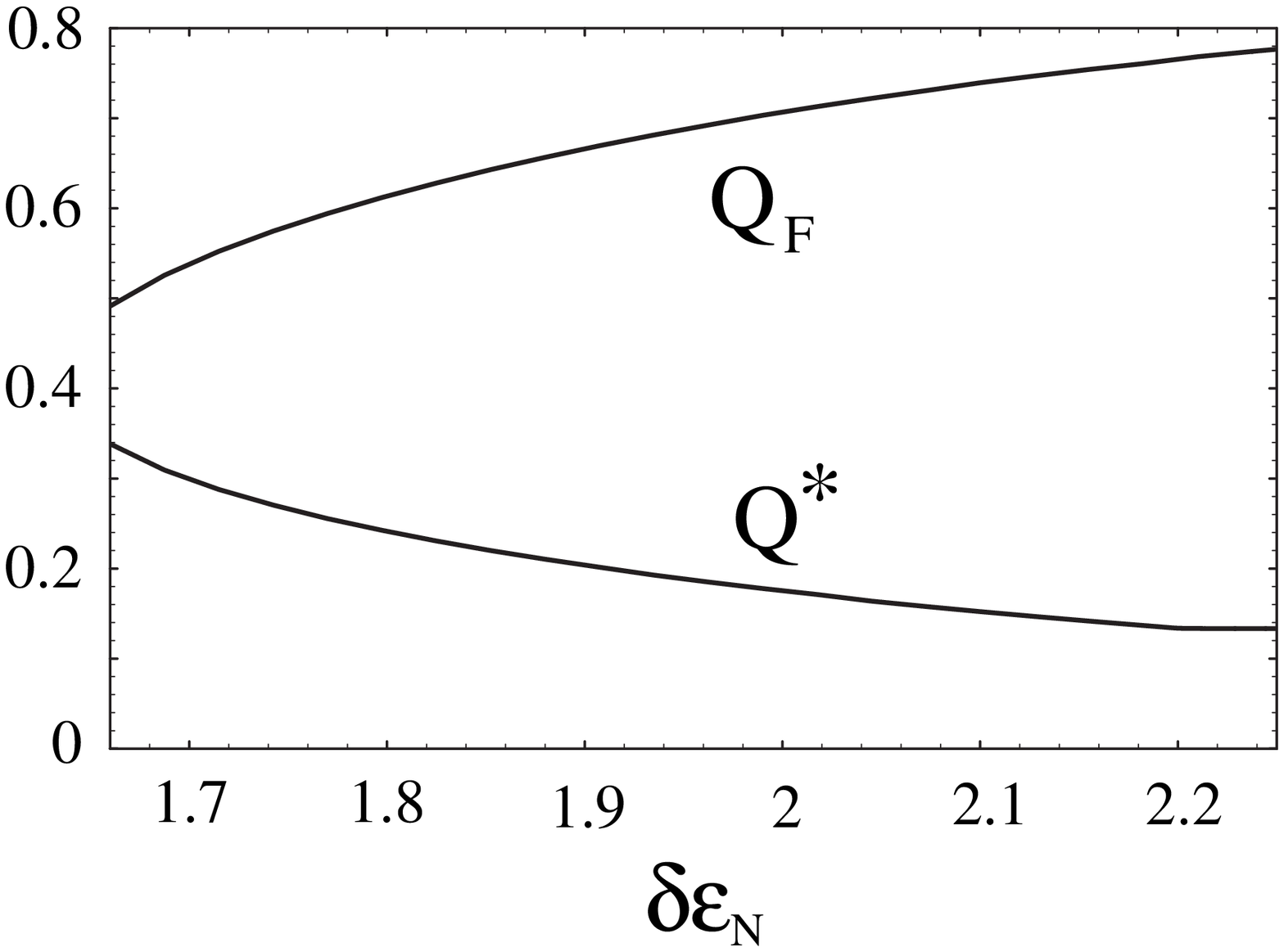,width=0.8\textwidth,angle=0}}
\centerline{\psfig{figure=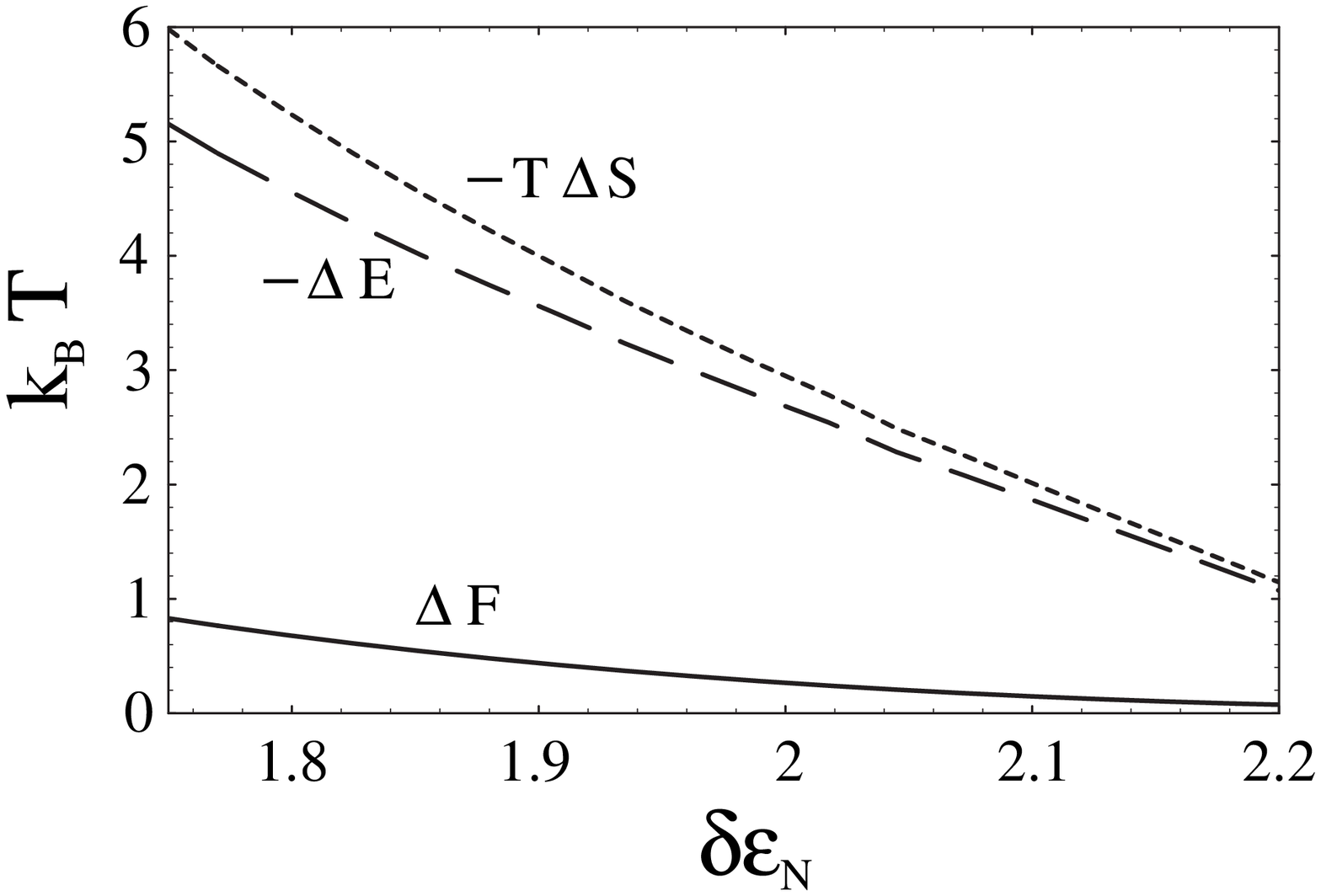,
	width=0.8\textwidth,angle=0}}
\caption{}
\label{fig:FqstVsEn}
\end{figure}

\begin{figure}
\centerline{\psfig{figure=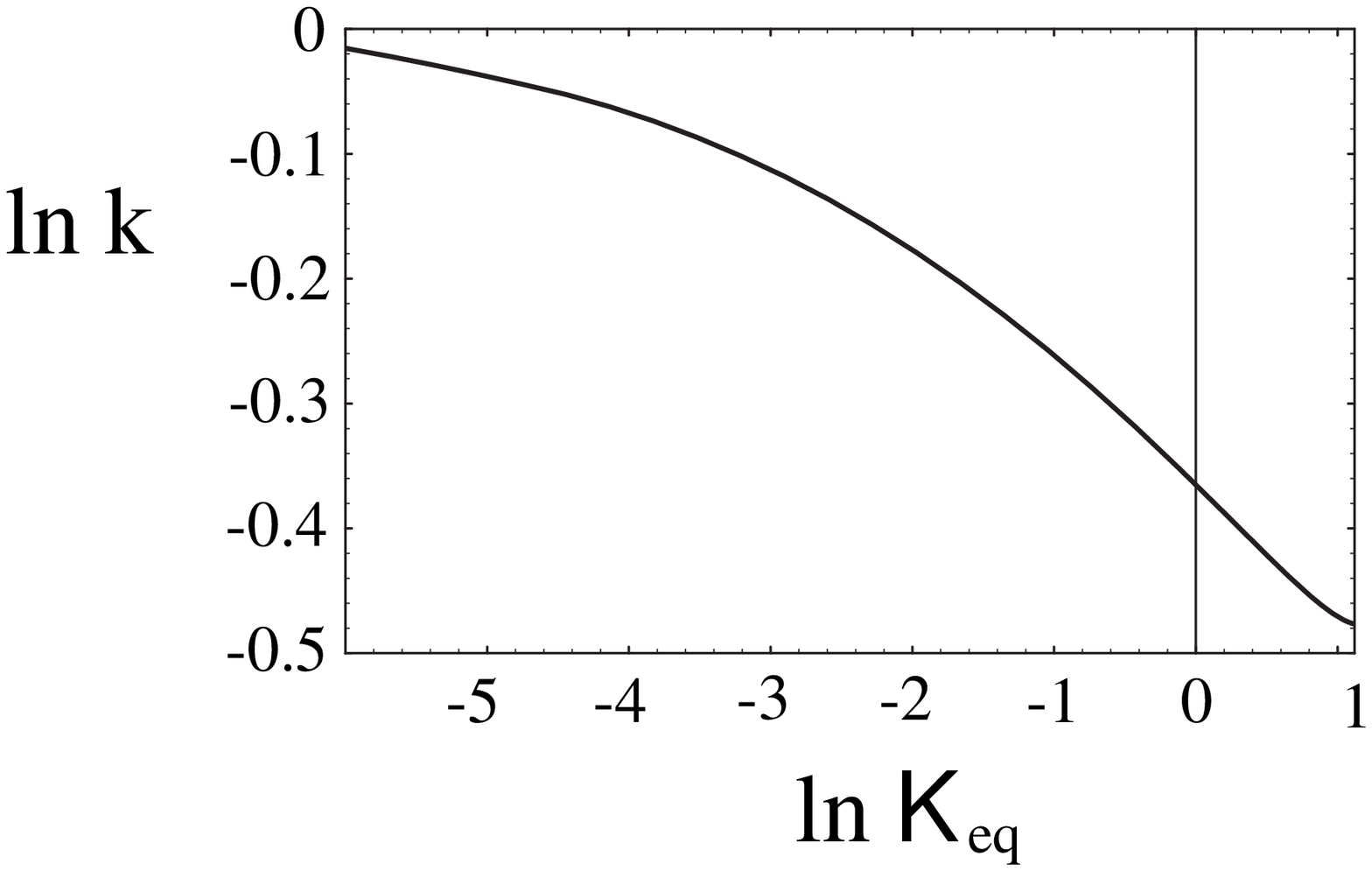,width=0.8\textwidth,angle=0}}
\centerline{\psfig{figure=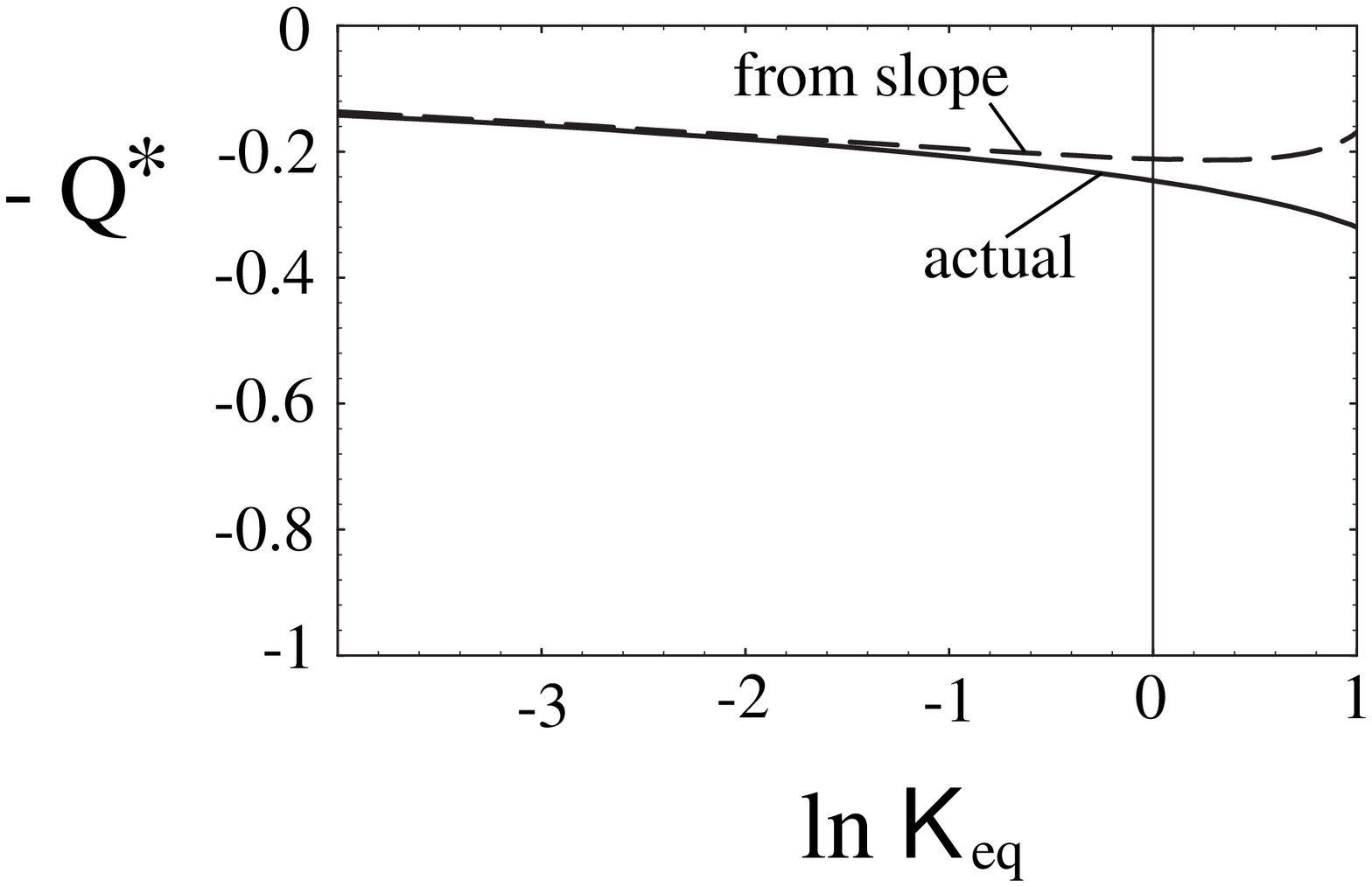,width=0.8\textwidth,angle=0}}
\caption{}
\label{fig:lnlnK}
\end{figure}

\begin{figure}
\centerline{\psfig{figure=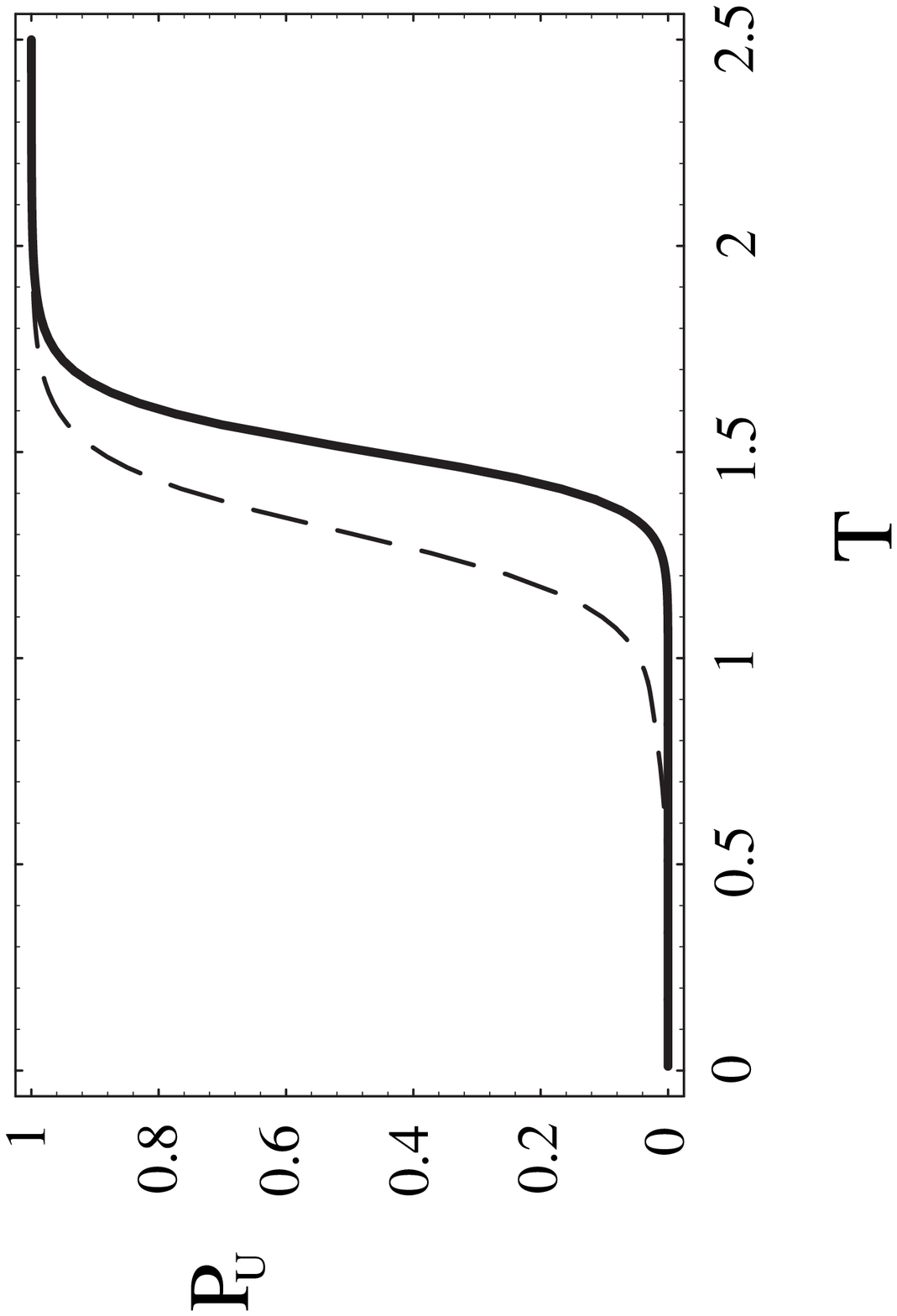,width=0.8\textwidth,angle=0}}
\caption{}
\label{fig:pdenat2}
\end{figure}

\begin{itemize}

\item [Figure~\ref{fig:FvsQ3and12}.]
	{\bf (A)}  Free energy per monomer $F/N$ for a $27$-mer, 
	in units of $\kboltz T_{\mbox{\tiny F}}$
	as a function of $Q$, at constant density $\eta = 1$,
	with $s_o = 0.8$, for protein-like 
	energetic parameters ($\delen,\varepsilon$) = ($-2.28 , 1.55$).
	For these parameters $T_{\mbox{\tiny F}} \approx 
	\left|\delen\right|$.
	For illustrative purposes, two values of $m$-body interactions are
	chosen:  
	({\bf solid line}) Pure $3$-body interactions.
	({\bf dashed line}) Pure $12$-body interactions.
	Note the trends in height and position of the barrier, and 
	note how in the $m=12$ case the free energy curve is essentially
	$-T$ times the entropy curve $s(Q)$ until $Q$ is large.
	\newline	{\bf (B)}
	Position of the transition state ensemble $\qst$ along the reaction
	coordinate $Q$ as a function of the explicit cooperativity in 
	pure $m$-body forces, $m$.
	The fact that the asymptotic limit $Q_{max}$ is less than one 
	is due to the finite size of the system, so that $Q$, the 
	fraction of native contacts, is not
	a continuous parameter.
	\newline	{\bf (C)}
	The free energy barrier height $\Delta F$ in units of $\kboltz \tf$ 
	as a function of the explicit cooperativity of the $m$-body force,
	$m$.  The barrier height rises to the limit of $S(Q=0)$ as 
	$m \rightarrow \infty$, when it becomes completely entropic. 
	Also shown are the
	energetic (dashed) and entropic (solid) contributions to the barrier.
\item [Figure~\ref{fig:corehalo}.]
	A model of the partially native protein 
	can be pictured as a frozen native core surrounded by a halo of 
	non-native polymer of variable density.
\item [Figure~\ref{fig:tgQZ}.]
	The folding temperature $\tf$ and glass transition temperature $\tg$
	as a function of the fraction of native contacts $Q$ and the total 
	contacts per monomer $\zbar$.
	The folding temperature is above the glass temperature (\ref{eq:tgs})
	for essentially all values of $Q$ and $\zbar$, for protein-like 
	energetic parameters. Here $\varepsilon \approx 1$ and $\delen = -1.6$.
\item [Figure~\ref{fig:data}.]
	{\bf (A)}
	The free energy 
	vs. $Q$ and $\zbar$, at the folding temperature $\tf$, 
	from simulations.~\cite{owls}
	\newline  {\bf (B)}
	Free energy surface at $\tf$ for the $27$-mer, obtained 
	from eq. (\ref{eq:fqz}) with the 
	parameters $\Evect = ( 0.9 , -2.8 , -1.6 , 1.51 )$.
	The surface has a double well structure with a transition state 
	ensemble at $\qst \cong 0.5$.
\item [Figure~\ref{fig:q3body}.]
	{\bf (A)}
	Position of the barrier $\qst$ and position of the folded free energy 
	minimum $\qf$ as a function of the three body coefficient $\alpha$.
	\newline {\bf (B)}
	Free energy barrier $\Delta F = F(\qst,\zst) - F(\qmg,\zmg)$ 
	in units of $\kboltz \tf$, 
	and its energetic and entropic contributions, for the $27$-mer,
	as a function of $\alpha$. There are two values of $\alpha$ where 
	the barrier is completely
	entropic, which define a region where the transition ensemble has a 
	lower average energy than the molten globule.
\item [Figure~\ref{fig:tNplot}.]
	{\bf (A)}
	The folding temperature $\tf$ is an increasing function of 
	polymer sequence length $N$.
	\newline  {\bf (B)}
	Position of the barrier $\qst$ as a function of sequence length $N$.
	The solid line is the theory as determined by eq. (\ref{eq:fqz}), and
	the points marked by polygons are experimental results (see text).
	\newline  {\bf (C)}
	Free energy barrier height $\Delta F$ in units of $\kboltz \tf$,
	as a function of sequence length $N$,
	along with its energetic and entropic contributions.
\item [Figure~\ref{fig:FqstVsEn}.]
	{\bf (A)}  Two plots of the free energy vs. $Q$ for the $27$-mer with 
	$\varepsilon=0.9$ and $T = 1.5$, at fixed density $\eta=0.95$. 
	The upper curve
	is the free energy when the stability gap $\delen = 1.75$, and 
	$\delen = 2.1$ in the lower curve.  From the figure we can see 
	that as $\delen$ increases and the folding becomes downhill,
	 the barrier shifts to lower $Q$ and
	decreases in height.
	\newline {\bf (B)}  The positions of the barrier $\qst$ 
	and folded state $\qf$
	as a function of energy gap $\delen$, for the system described
	in (A), where $T=1.5$ and $\varepsilon = 0.9$. \newline
	{\bf (C)} Free energy barrier in units of $\kboltz T$ vs.
	stability gap $\delen$.
 	The short dashed line is the entropic contribution to the 
	barrier, and the long dashed line is minus the energetic contribution.
\item [Figure~\ref{fig:lnlnK}.]
	{\bf (A)} plot of the logarithm of the 
	folding rate vs. the logarithm of the unfolding 
	equilibrium constant.
	\newline {\bf (B)}
	Reading the slope from a (A) gives a measure of the position 
	of the barrier $\qst$.
	Also plotted is the actual value of $\qst$ directly calculated 
	from the free energy curves.  The values compare well for most values
	of the gap where the free energy has a double-well structure.
\item [Figure~\ref{fig:pdenat2}.]
	{\bf (solid line)} Probability to be in the unfolded state 
	(eq. [\ref{eq:pd}])
	 as a function of 
	temperature for the $27$-mer with
	energetic parameters $(\varepsilon,\delen,\tf) = (0.9,-1.8,1.51)$.
	\newline
	{\bf (dashed line)} Same probability for a protein with a rougher 
	energy landscape, which has a
	lower folding temperature and somewhat broader denaturation curve,
	$(\varepsilon,\delen,\tf) = (1.3,-2.0,1.3)$.

\end{itemize}
\newpage

\begin{thebibliography}{99}
\bibitem{snote} Unless otherwise indicated, all entropies have
	dimensions of Boltzmann's constant $\kboltz$.
\bibitem{kboltznote} Unless otherwise indicated, all temperatures 
	have units of energy, with Boltzmann's constant 
	$\kboltz$ included in the definition of $T$.
\end{thebibliography}
\end{document}